\documentclass[twocolumn,aps,prb,reprint,twocolumn,superscriptaddress,amsmath,amssymb]{revtex4-1}
\usepackage[latin9]{inputenc}
\usepackage{color}
\usepackage{float}
\usepackage{amsmath}
\usepackage{graphicx}
\usepackage[unicode=true,pdfusetitle,
 bookmarks=false,
 breaklinks=false,pdfborder={0 0 1},backref=false,colorlinks=false]
 {hyperref}

\makeatletter

\providecommand{\tabularnewline}{\\}

\usepackage{bbold}
\makeatother

\begin{document}
\title{A Generalized Open Quantum System Approach for the Electron Paramagnetic
Resonance of Magnetic Atoms }
\author{Gal Shavit}
\affiliation{Raymond and Beverly Sackler School of Physics and Astronomy, Tel Aviv
University, Tel Aviv 6997801, Israel}
\author{Baruch Horovitz}
\affiliation{Department of Physics, Ben-Gurion University of the Negev, Beer Sheva
84105, Israel}
\author{Moshe Goldstein}
\affiliation{Raymond and Beverly Sackler School of Physics and Astronomy, Tel Aviv
University, Tel Aviv 6997801, Israel}
\begin{abstract}
A recent experimental breakthrough allowed to probe electronic parametric
resonance of a single magnetic atom in a scanning tunneling microscopy
(STM) setup. The results present intriguing features, such as an asymmetric
lineshape and unusually large ratio of the decoherence and decay rates,
which defy standard approaches using the conventional Bloch equations.
To address these issues we employ novel generalized Bloch equations,
together with proper microscopic modeling of the magnetic adatom,
and show how all the experimental features can be naturally accounted
for. The proposed approach may also be useful in treating any future
similar experiments, as well as next generation hybrid quantum devices. 
\end{abstract}
\maketitle

\section{Introduction}

Electron paramagnetic resonance (EPR) experiments have been a powerful
tool in studying the properties of different paramagnetic materials
by probing the spin of unpaired electrons for several decades \cite{abragam1970electron}.
Recently, the possibility of single spin resolution in EPR detection
has been realized by utilizing STM to measure the tunneling conductance
through a magnetic impurity \cite{pumpprobe,STMexperiment,Willke2018,CHOI_ESR_STM,Choi2017,WillkeNew},
where a spin-polarized STM tip is used both as the EPR pump and probe.
A different type of EPR-STM phenomena was realized by using a non-polarized
tip \cite{StmExample,BaruchEndor}.

Focusing here on experiments with polarized tips, the results of these
experiments pose several difficulties. First, a conspicuous asymmetry
in the resonance lineshape, with the signal even dropping below its
asymptotic value; this was previously attributed to phenomenological
Fano interferences \cite{Willke2018,JETPfano}. Second, a $T_{1}$
relaxation time which is about three orders of magnitude longer than
the decoherence time $T_{2}$. As we will show, these features (and
others to be detailed below) can naturally be explained as intrinsic
effects provided we: (a) go beyond the traditional Bloch equation
employed in these works, and use a new generalized quantum master
equation; (b) account for the fact that the two level system addressed
by the EPR excitation is a part of a more complex energy manifold
of the adatom; (c) derive relaxation rates from the spin-electrode
couplings.

The paper is organized as follows. In Sec. \ref{modelSec} we present
the theoretical model for the experimental system, as well as its
mapping to an effective open quantum two-level system (TLS). We then
show how a measurement of the tunneling current reveals the steady
state polarization of the TLS in Sec. \ref{currentToPolarizationSec}.
The flaws in treating the system using traditional approaches are
pointed out in Sec. \ref{BlochFailure}, leading to an introduction
of our novel generalized approach in Sec. \ref{generalizedApproach}.
After showing that our proposed treatment can account for some of
the experimental observations, in Sec. \ref{HighV} we consider the
higher voltage regimes, where higher energy levels play a role, and
show that our approach captures the unique features observed in this
regime. We summarize our findings in Sec. \ref{cocncluze}. We give
additional technical details regarding the derivation of our generalized
approach in Appendix \ref{4genME}, and expand on the calculation
of relaxation and decoherence rates in Appendix \ref{ratesApp}.

\section{Model}

\label{modelSec}

\begin{figure}
\begin{centering}
\includegraphics[scale=0.4]{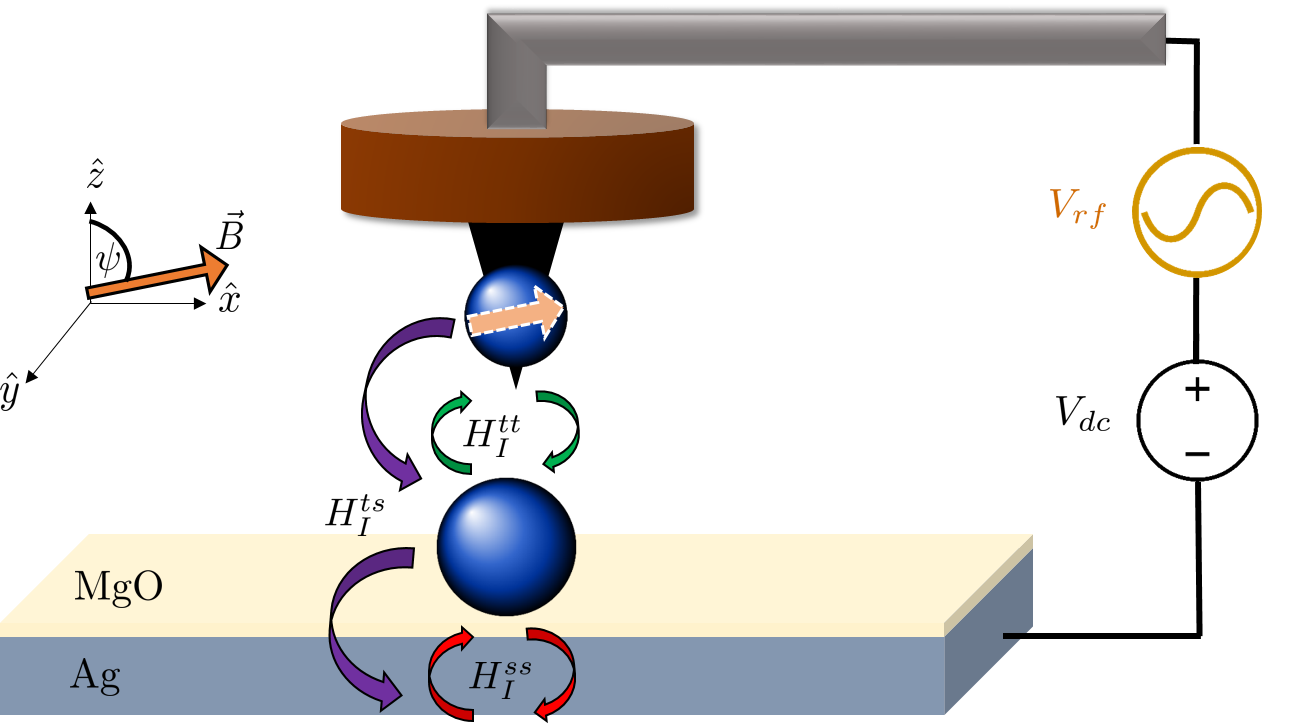} 
\par\end{centering}
\caption{\label{fig:Schematic-description-of} Experimental setup of the EPR-STM
experiment \cite{STMexperiment,Willke2018}. The iron adatom is placed
on an MgO substrate below an STM tip. Applying an appropriate magnetic
field $\vec{B}$ effectively turns the atom into a TLS in the working
dc voltage regime and polarizes the STM tip in the field direction.
An additional rf voltage is applied between the tip and the substrate
to measure the electron paramagnetic resonance in the atom. Also shown
is a schematic depiction of the exchange interaction processes between
the adatom and the bath electrons. The strength of the exchange interaction
for electrons hopping from tip to the adatom and back (green) is $J_{t}$,
for electrons hopping from the substrate and back (red) is $J_{s}$,
and for electrons tunneling from the tip to the substrate (or the
other way around) through the adatom (purple) is $\sqrt{J_{t}J_{s}}$.}
\end{figure}

To make the discussion concrete we concentrate on the system studied
in Ref. \cite{STMexperiment} (see Fig. \ref{fig:Schematic-description-of}).
There, single iron atoms were placed on a monolayer magnesium oxide
(MgO) film, isolating the atoms from a bulk silver substrate. A spin-polarized
STM tip was positioned above the iron adatom, with the direction of
its polarization determined by the applied magnetic field, which is
at an angle $\psi$ to the axis perpendicular to the MgO plane, whose
value was close to $90^{\circ}$ (field nearly parallel to the substrate).
The magnetic atom placed on the substrate (assumed to be in the $\mbox{d\ensuremath{^{6}}}$
electronical configuration in the lowest Hund's term, with $L=S=2$)
is well described by the ligand-field Hamiltonian \cite{STMfeOnMgO}

\begin{equation}
H_{{\rm lf}}=DL_{z}^{2}+EL_{z}^{4}+F_{0}\left(L_{+}^{4}+L_{-}^{4}\right)+\lambda\vec{L}\cdot\vec{S}+\mu_{B}\left(\vec{L}+2\vec{S}\right)\cdot\vec{B},\label{eq:LFH}
\end{equation}

\noindent with the applied magnetic field $\vec{B}$ and the Bohr
magneton $\mu_{B}$. The values used for the parameters in $H_{{\rm lf}}$
are given in Table \ref{tab:ligandParams}. This Hamiltonian includes
all terms allowed by the four-fold symmetry of the Fe bound to the
MgO layer. A finite magnetic field component in the direction perpendicular
to the MgO substrate acts as a Zeeman field, splitting the lowest
energy state of the atom into an effective TLS, polarized in its spin
component, and isolated from the rest of the spectrum by a gap of
roughly $\sim14$ meV \cite{STMexperiment}. A dc voltage $V_{{\rm dc}}=5$
mV was set between the tip and the substrate, allowing tunneling of
electrons between the tip and the bulk substrate through the adatom.
Additionally, an rf voltage was introduced, driving coherent transitions
of the TLS.

\begin{table}[H]
\begin{centering}
\begin{tabular}{|c|c|}
\hline 
\textbf{\small{}{}Parameter}{\small{} } & \textbf{\small{}{}Approximate value}\tabularnewline
\hline 
\hline 
{\small{}{}$D$}  & {\small{}{}$-433$ meV}\tabularnewline
\hline 
{\small{}{}$E$}  & {\small{}{}$0$ meV}\tabularnewline
\hline 
{\small{}{}$F_{0}$}  & {\small{}{}$2.19$ meV}\tabularnewline
\hline 
{\small{}{}$\lambda$}  & {\small{}{}$-12.6$ meV}\tabularnewline
\hline 
\end{tabular}
\par\end{centering}
\caption{\label{tab:ligandParams} Approximate values for the free parameters
in the ligand-field Hamiltonian (Eq. \eqref{eq:LFH}), see Supplementary
Material for \cite{STMexperiment}.}
\end{table}

\noindent Projecting \eqref{eq:LFH} into its two lowest levels and
including the periodic drive and the coupling of the atom to the tip
and substrate electrons, one finds the total Hamiltonian 
\begin{equation}
H=H_{S}+H_{D}+H_{I}+H_{B},\label{eq:GenHamiltonian}
\end{equation}
with $H_{S}$, $H_{D}$, $H_{I}$ and $H_{B}$ representing the system,
periodic driving, interaction of the TLS with the electronic bath
(phononic dissipation is neglected as a result of the low temperature,
$T\approx0.6\,K$) and bath Hamiltonians respectively, and are given
by \begin{subequations} 
\begin{equation}
H_{S}=-\frac{1}{2}\hbar\omega_{0}\sigma_{z},
\end{equation}
\begin{equation}
H_{D}=\hbar\Omega\cos\left(\omega_{d}t\right)\sigma_{x},
\end{equation}
\begin{equation}
H_{I}=H_{I}^{ss}+H_{I}^{tt}+H_{I}^{ts},
\end{equation}
\begin{equation}
H_{B}=\sum_{k,\sigma,\ell}\left(\epsilon_{k\sigma\ell}c_{k\sigma\ell}^{\dagger}c_{k\sigma\ell}+T_{0}c_{k\sigma\ell}^{\dagger}c_{k\sigma\bar{\ell}}\right),
\end{equation}
\end{subequations}where $\vec{\sigma}$ are the Pauli matrices in
the TLS subspace, $\hbar\omega_{0}$ is the two-level energy separation,
$\ensuremath{\Omega\propto V_{{\rm rf}}}$ is the driving amplitude,
$\omega_{d}\equiv\omega_{0}+\delta\omega$ is the driving frequency
(typically $\Omega=1-20$ MHz, $\omega_{0},\omega_{d}\sim25$ GHz.
The driving $\Omega$ and detuning $\delta\omega$ combine to give
the generalized Rabi frequency $\omega\equiv\sqrt{\Omega^{2}+\delta\omega^{2}}$),
$T_{0}$ is the tip-substrate tunneling amplitude, and $c_{k\sigma\ell}$
is an electronic annihilation operator with momentum $k$ and spin
$\sigma$ in reservoir $\ell$ (either the tip or the substrate, with
$\bar{\ell}$ the reservoir opposing $\ell$). The three different
components of the interaction Hamiltonian $H_{I}$ describe different
hopping processes of electrons: tip-atom-tip ($tt$), substrate-atom-substrate
($ss$), and tip-atom-substrate ($ts$), see Fig. \ref{fig:Schematic-description-of}.
They are of the form 
\begin{equation}
H_{I}^{\ell\ell'}=\sqrt{J_{\ell}J_{\ell'}}\vec{S}\cdot\vec{S}_{\ell\ell'},\label{eq:IntH}
\end{equation}
with $\vec{S}_{\ell\ell'}=\sum_{k,k'}\left(c_{k\sigma\ell}^{\dagger}\vec{\sigma}_{\sigma\sigma'}c_{k'\sigma'\ell'}+\mathrm{h.c.}\right)$.
$\vec{S}$ is the \textit{physical} spin operator of the magnetic
atom, while $\sigma$, whenever next to $c,c^{\dagger}$ operators,
operates within the spin space of the tunneling electrons.

\subsection{Mapping the interaction Hamiltonian into an open TLS}

The Hamiltonian Eq. \eqref{eq:LFH} can be diagonalized in the $5\times5$
Hilbert space spanning the different orbital momentum and spin states.
We shall henceforth focus on the two lowest energy levels of $H_{{\rm lf}}$,
denoted by $\left|0\right\rangle $ and $\left|1\right\rangle $.
Any operator $\hat{O}$ can be projected onto the subspace spanned
by the TLS, 
\begin{eqnarray}
\hat{O} & = & \frac{1}{2}\sigma_{x}\left(O_{10}+O_{01}\right)+\frac{i}{2}\sigma_{y}\left(O_{10}-O_{01}\right)\nonumber \\
 &  & +\frac{1}{2}\sigma_{z}\left(O_{00}+O_{11}\right)+\frac{1}{2}\left(O_{00}+O_{11}\right),
\end{eqnarray}
with the matrix elements $O_{ij}\equiv\left\langle i\right|\hat{O}\left|j\right\rangle $.
In this manner, we can map all spin operators in the two-level subspace
using the general notation 
\begin{equation}
\begin{pmatrix}S_{x}\\
S_{y}\\
S_{z}
\end{pmatrix}=\begin{pmatrix}\alpha_{xx} & \alpha_{xy} & \alpha_{xz}\\
\alpha_{yx} & \alpha_{yy} & \alpha_{yz}\\
\alpha_{zx} & \alpha_{zy} & \alpha_{zz}
\end{pmatrix}\begin{pmatrix}\sigma_{x}\\
\sigma_{y}\\
\sigma_{z}
\end{pmatrix}.
\end{equation}
Using the parameters in Table \ref{tab:ligandParams}, we evaluate
the spin projection matrix, 
\begin{equation}
\begin{pmatrix}\alpha_{xx} & \alpha_{xy} & \alpha_{xz}\\
\alpha_{yx} & \alpha_{yy} & \alpha_{yz}\\
\alpha_{zx} & \alpha_{zy} & \alpha_{zz}
\end{pmatrix}\approx-2\cdot10^{-4}\begin{pmatrix}2.4 & 0 & -1\\
0 & 2.4 & 0\\
7.1 & 0 & 10^{4}
\end{pmatrix}.\label{eq:alphaMatrix-1}
\end{equation}
Importantly, $\alpha_{zz}$ is by far the most dominant matrix element,
and we find the approximation $\alpha_{xx}\approx\alpha_{yy}\equiv\alpha_{\perp}$
well justified.

The quantization axis of the spin in the interaction Hamiltonian \eqref{eq:IntH},
the $\hat{z}$ axis, is directed along the ``lab'' $\hat{z}$ direction,
normal to the substrate (which is also the easy axis of the adatom
deposited on the MgO layer). We perform a rotation on the electronic
operators such that the new spin axis lies in the direction of the
STM tip spin-polarization (similarly to \cite{STMextendedTheory})
\begin{subequations} 
\begin{equation}
c_{k\uparrow\ell}\rightarrow\cos\frac{\psi}{2}c_{k\uparrow\ell}-\sin\frac{\psi}{2}c_{k\downarrow\ell},
\end{equation}
\begin{equation}
c_{k\downarrow\ell}\rightarrow\sin\frac{\psi}{2}c_{k\uparrow\ell}+\cos\frac{\psi}{2}c_{k\downarrow\ell}.
\end{equation}
\end{subequations} After this rotation, we write the interaction
Hamiltonian as 
\begin{equation}
H_{I}^{\ell\ell'}\equiv\sqrt{J_{\ell}J_{\ell'}}\left(\sigma_{+}\hat{\Sigma}_{+}+\sigma_{-}\hat{\Sigma}_{-}+\sigma_{z}\hat{\Sigma}_{z}\right),\label{eq:GeneralOpenSystem}
\end{equation}
with the electronic operators $\hat{\Sigma}_{i}$,

\begin{align}
\hat{\Sigma}_{-} & =\left(\alpha_{zx}\cos\psi+\frac{\alpha_{\perp}}{2}\sin\psi\right)\times\nonumber \\
 & \sum_{k,k',\sigma}\sigma\left(c_{k\sigma\ell}^{\dagger}c_{k'\sigma\ell'}+c_{k'\sigma\ell'}^{\dagger}c_{k\sigma\ell}\right)\nonumber \\
 & +\left(\frac{\alpha_{\perp}}{2}\left(\cos\psi+1\right)-\alpha_{zx}\sin\psi\right)\times\nonumber \\
 & \sum_{k,k'}\left(c_{k\downarrow\ell}^{\dagger}c_{k'\uparrow\ell'}+c_{k'\downarrow\ell'}^{\dagger}c_{k\uparrow\ell}\right)\nonumber \\
 & +\left(\frac{\alpha_{\perp}}{2}\left(\cos\psi-1\right)-\alpha_{zx}\sin\psi\right)\nonumber \\
 & \sum_{k,k'}\left(c_{k\uparrow\ell}^{\dagger}c_{k'\downarrow\ell'}+c_{k'\uparrow\ell'}^{\dagger}c_{k\downarrow\ell}\right),
\end{align}
\begin{align}
\hat{\Sigma}_{z} & =\left(\alpha_{zz}\cos\psi+\frac{1}{2}\alpha_{xz}\sin\psi\right)\sum_{k,k',\sigma}\sigma c_{k\sigma\ell}^{\dagger}c_{k'\sigma\ell'}\nonumber \\
 & +\left(\frac{1}{2}\alpha_{xz}\cos\psi-\alpha_{zz}\sin\psi\right)\sum_{k,k',\sigma}c_{k\sigma\ell}^{\dagger}c_{k'\bar{\sigma}\ell'}+\mathrm{h.c.}\,,
\end{align}
and $\hat{\Sigma}_{+}=\left(\hat{\Sigma}_{-}\right)^{\dagger}$.

\section{Relating the tunneling current to the TLS polarization}

\label{currentToPolarizationSec}

We consider the tunneling current due to interactions with the reservoirs,
following mainly the treatment in \cite{STMextendedTheory}. The labels
$\ell$, $\ell'$ refer here to the tip and substrate, respectively.
The current operator may be determined by examining the change in
total charge over time in one of the leads, 
\begin{equation}
I\left(t\right)=ie\left[\sum_{k,\sigma}c_{k\sigma t}^{\dagger}c_{k\sigma t},H\right]=ie\left(C-C^{\dagger}\right),
\end{equation}
with 
\begin{equation}
C\equiv T_{0}\sum_{k,k',\sigma}c_{k\sigma t}^{\dagger}c_{k'\sigma s}e^{ieVt}+J_{ts}\sum_{k,k',\sigma,\sigma'}c_{k\sigma t}^{\dagger}\vec{\sigma}_{\sigma\sigma'}\cdot\vec{S}c_{k'\sigma's}e^{ieVt},
\end{equation}
and $J_{ts}=\sqrt{J_{t}J_{s}}$. We can expand the average of the
current operator to lowest (first) nonvanishing order in the tunneling
Hamiltonian, using the relation $\left\langle \hat{O}\left(t\right)\right\rangle =-i\int_{-\infty}^{t}dt'\left\langle \left[\hat{O}\left(t\right),H_{T}\left(t'\right)\right]\right\rangle _{0}$,
where the average inside the integral is taken in the zero tunneling
state. We find 
\begin{equation}
\left\langle I\left(t\right)\right\rangle =e\int_{-\infty}^{t}dt'\left\langle \left[C\left(t'\right),C^{\dagger}\left(t\right)\right]\right\rangle _{0}+\mathrm{c.c.}\label{eq:Ikeldysh}
\end{equation}

The expression in Eq. \eqref{eq:Ikeldysh} can be divided into three
individual contributions, proportional, respectively, to $T_{0}^{2}$,
$J_{ts}^{2}$ and $T_{0}J_{ts}$. The first contribution corresponds
to background current, which is unaffected by the adatom spin. The
second term will be significantly weaker compared to the last one
due to the fact that typically $\frac{J_{ts}}{T_{0}}\sim0.1$ (the
ratio between the spin exchange energy and the spin-independent tunneling
amplitude \cite{STMextendedTheory}). Hence, we shall focus on the
$T_{0}J_{ts}$ term, which we denote by $I_{T_{0}J}$. Neglecting
any scattering between the different momentum/spin channels in the
$\left\langle ...\right\rangle _{0}$ average, and using the Fermi-Dirac
distribution $f\left(\epsilon_{k}\right)$, we find 
\begin{eqnarray}
\left\langle I_{T_{0}J}\left(t\right)\right\rangle  & = & eT_{0}J_{ts}\sum_{k,k'\sigma}\sigma_{\sigma\sigma}^{z}\int_{-\infty}^{t}dt'e^{i\left[eV+\epsilon_{k\sigma}-\epsilon_{k'\sigma}\right]\left(t'-t\right)}\times\nonumber \\
 &  & \left[f_{\ell'\sigma}\left(k\right)\left(1-f_{\ell\sigma}\left(k'\right)\right)\left\langle S^{z}\left(t'\right)\right\rangle \right.\nonumber \\
 &  & \left.-f_{\ell\sigma}\left(k\right)\left(1-f_{\ell'\sigma}\left(k'\right)\right)\left\langle S^{z}\left(t\right)\right\rangle \right]+\mathrm{c.c.}
\end{eqnarray}
We set the local densities of states $\nu_{\ell'\sigma}=\frac{\nu_{t}}{2}\left(1+\sigma p\right),$
$\nu_{\ell\sigma}=\frac{\nu_{s}}{2}$, which is possible since we
choose here the $\hat{z}$ direction to \textit{be the tip spin-polarization
axis}, denoted as $\hat{p}$ in the following. Moving to a summation
over energies instead of momenta, and taking the long-time limit such
that $\left\langle S_{\hat{p}}\right\rangle $ reaches its steady
state, 
\begin{eqnarray}
\left\langle I_{T_{0}J}\right\rangle  & = & eT_{0}J_{ts}\frac{\nu_{s}\nu_{t}}{2}p\left\langle S_{\hat{p}}\right\rangle \int_{-\infty}^{\infty}d\tau e^{i\left(eV+\epsilon-\epsilon'\right)\tau}\times\nonumber \\
 &  & \int_{-\infty}^{\infty}d\epsilon'\int_{-\infty}^{\infty}d\epsilon\times\nonumber \\
 &  & \left[f\left(\epsilon\right)\left(1-f\left(\epsilon'\right)\right)-f\left(\epsilon'\right)\left(1-f\left(\epsilon\right)\right)\right]+\mathrm{c.c.}
\end{eqnarray}
Performing the integration over $\tau$ will result in a delta function.
Assuming $eV\gg k_{B}T$ (in the experiment $eV=5$--$60$ meV and
$k_{B}T\approx50$ $\mu$eV), we find 
\begin{equation}
\left\langle I_{T_{0}J}\right\rangle \approx\pi e^{2}VT_{0}J_{ts}\nu_{s}\nu_{t}p\left\langle S_{\hat{p}}\right\rangle ,
\end{equation}

\noindent i.e., the spin-dependent contribution to the current is
proportional to the steady-state spin polarization of the intermediary
adatom \textit{in the direction of the tip polarization}.

The physical spin polarization can be expressed in terms of TLS expectation
values, 
\begin{align}
\left\langle S_{\hat{p}}\right\rangle  & =\left(\alpha_{xz}\sin\psi+\alpha_{zz}\cos\psi\right)\left\langle \sigma_{z}\right\rangle \nonumber \\
 & +\left(\alpha_{\perp}\sin\psi+\alpha_{zx}\cos\psi\right)\left\langle \sigma_{x}\right\rangle .
\end{align}
Since $\alpha_{zz}$ overwhelmingly dominates the other matrix elements,
and taking into account $\psi\approx80^{\circ}-88^{\text{\ensuremath{\circ}}}$
in these experiments, we conclude that 
\begin{equation}
\left\langle I_{T_{0}J}\right\rangle \propto\alpha_{zz}\left\langle \sigma_{z}\right\rangle ,
\end{equation}
so that the tunneling current is a direct measurement of the TLS polarization.
Sweeping the driving frequency and measuring the change in current
(compared to the non-driven $\Omega=0$ case), resonant lineshapes
may be observed, with the resonant frequency corresponding to the
energy separation of the TLS.

\section{Failure of standard Bloch equations}

\label{BlochFailure}

Ref. \cite{STMexperiment} analyzed the TLS dynamics using the Bloch
equations {[}Supplementary Material for \cite{STMexperiment}, Eq.
(S1){]}. Writing the adatom density matrix in the form $\rho\equiv\frac{1}{2}+\left(n-\frac{1}{2}\right)\sigma_{z}+\alpha^{*}\sigma_{-}+\alpha\sigma_{+}$,
the equations may be written in the frame rotating with $\omega_{d}$
as\begin{subequations} 
\begin{eqnarray}
\frac{d}{dt}n & = & -n\left(\Gamma_{\downarrow}+\Gamma_{\uparrow}\right)+\Gamma_{\uparrow}-i\Omega\frac{\alpha-\alpha^{*}}{2},\label{eq:BlochN}
\end{eqnarray}
\begin{eqnarray}
\frac{d}{dt}\alpha & = & -\alpha\left(\frac{\Gamma_{\downarrow}+\Gamma_{\uparrow}}{2}+2\Gamma^{z}+i\delta\omega\right)-i\Omega\left(n-\frac{1}{2}\right),\,\,\,\,\,\,\,\,\,\label{eq:BlochA}
\end{eqnarray}
\end{subequations} with $\Gamma_{\downarrow}$, $\Gamma_{\uparrow}$
and $\Gamma^{z}$ the relaxation, excitation, and pure dephasing rates,
which are given by ($a=-1,1$ corresponding to $\downarrow$, $\uparrow$,
respectively) 
\begin{eqnarray}
\Gamma_{a} & \equiv & \frac{1}{2}\mathrm{Re}\left\{ \int_{0}^{\infty}d\tau e^{-ia\omega_{d}\tau}\mathrm{Tr}_{B}\left\{ \rho_{B}\hat{\Sigma}_{a}\left(\tau\right)\hat{\Sigma}_{-a}\left(0\right)\right\} \right\} ,\,\,\,\,\,\,\,\label{eq:ks_by_a}
\end{eqnarray}
\begin{eqnarray}
\Gamma^{z} & \equiv & \frac{1}{2}\mathrm{Re}\left\{ \int_{0}^{\infty}d\tau\mathrm{Tr}_{B}\left\{ \rho_{B}\hat{\Sigma}_{z}\left(\tau\right)\hat{\Sigma}_{z}\left(0\right)\right\} \right\} ,\label{eq:ks_by_a-1}
\end{eqnarray}
where $\mathrm{Tr}_{B}\left\{ \cdot\right\} $ is a trace over the
bath degrees of freedom (the reservoir electrons), and $\rho_{B}$
is the bath density matrix. The commonly used decay times are then
$\frac{1}{T_{1}}=\Gamma_{\downarrow}+\Gamma_{\uparrow}$ and $\frac{1}{T_{2}}=\frac{1}{2T_{1}}+2\Gamma^{z}\equiv\tilde{\Gamma}$,
where in the experiment Ref. \cite{STMexperiment} $T_{1}\sim100$
$\mu$sec, $T_{2}\sim100$ nsec. Eqs. \eqref{eq:BlochN}--\eqref{eq:BlochA}
are obtained by first deriving a master equation for the density matrix
in the absence of driving, and then adding the drive ``after the
fact'', such that it does not impact the dissipative terms. In particular,
it assumes relaxation towards the lab frame z axis, as if the drive
were absent. As our more general treatment will show, this \textit{lab
frame approach}, common mainly in atomic physics \cite{MollowRef4,AtomicRhobidium}
and quantum optics \cite{QuantumOptics},\textcolor{red}{{} }neglects
the difference between the values of the bath spectral functions at
frequencies $0,\pm\omega_{d}$ {[}Eqs. \eqref{eq:ks_by_a}--\eqref{eq:ks_by_a-1}{]},
and their values at $\pm\omega,\pm\omega_{d}\pm\omega$, respectively.
While one might expect these differences to be small for $\omega\ll T,V,\omega_{d}$,
which is the case here, we will show below that the importance of
some of these small differences is enhanced due to the non-equilibrium
nature of the system.

Solving \eqref{eq:BlochN}--\eqref{eq:BlochA} for the steady state
of the system and extracting $\left\langle \sigma_{z}\right\rangle =n-\frac{1}{2}$
results in 
\begin{equation}
\ensuremath{\left\langle \sigma_{z}^{\infty}\right\rangle {}_{\mathrm{Bloch}}}=\frac{\left(\Gamma_{\uparrow}-\Gamma_{\downarrow}\right)\left(\tilde{\Gamma}^{2}+\delta\omega^{2}\right)}{\left(\Gamma_{\downarrow}+\Gamma_{\uparrow}\right)\left(\tilde{\Gamma}^{2}+\delta\omega^{2}\right)+\tilde{\Gamma}\Omega^{2}},\label{eq:labframSz}
\end{equation}
which is \textit{even} in the detuning frequency $\delta\omega$,
and thus cannot reproduce the distinct asymmetric shape \textcolor{black}{observed
in many e}xperiments on these systems \cite{STMexperiment,Willke2018,WillkeNew,Choi2017,CHOI_ESR_STM}.
The asymmetric lineshape was attributed to an extrinsic effect, namely
th\textcolor{black}{e interplay between the precession of the tunnel
conductance and the rf voltage (Supplementary Materials to Refs. \cite{Willke2018,YANG2017}).
However, this phenomenological description does not contain a physical
reason for neither the conductance oscillations nor for the asymmetric
lineshape.}\textcolor{red}{{} }Moreover, the unusually high value
of the ratio $T_{1}/T_{2}$ found in these experiments still needs
to be accounted for. Below we show that both phenomena are \textit{intrinsic}
to the system.

One common possible alternative approach which includes modifications
to the dissipator stemming from the driving, may be obtained by diagonalizing
$H_{S}+H_{D}$ in a frame rotating with the driving frequency $\omega_{d}$,
and only then calculating the dissipative dynamics, now with a modified
system-bath interaction due to the driving. However, in order for
the master equation to be of Lindblad form \cite{lindbladSemigroups,GoriniGenerators},
this \textit{rotating frame approach} requires an additional ``secular
approximation'' with regards to the generalized Rabi frequency $\omega$,
i.e., that it is sufficiently greater than all the dissipative rates.
This approximation is inadequate in the experiments discussed here,
where $\omega T_{2}\lesssim1$.

\section{Generalized Bloch equations}

\label{generalizedApproach}

We develop and solve a novel \textit{generalized approach} by working
in the rotating frame but avoiding the customary secular approximation
\cite{REDFIELD19651} with regards to the low Rabi frequency, while
keeping the secular approximation only for the high frequencies $\omega_{0}$,$\omega_{d}$.
Our approach then covers the entire crossover range between $\omega=0$
(lab frame approach) and $\omega\gg\frac{1}{T_{1}},\frac{1}{T_{2}}$
(rotating frame approach). Crucially, the generalized approach keeps
the distinction between bath correlations calculated at frequencies
$\pm\omega$ and those at $0$, while neglecting the distinction between
correlations at $\pm\omega_{d}$ and $\pm\omega_{d}\pm\omega$. The
latter distinction is negligible since for $\omega\ll\omega_{d}$,
as we have explicitly checked. The distinction between frequencies
$\pm\omega$ and $0$ necessitates the introduction of $\Gamma_{\pm}^{z}$,
which are similar to $\Gamma_{z}$ {[}Eq. \eqref{eq:ks_by_a}{]} but
at frequencies $\pm\omega$, 
\begin{equation}
\Gamma_{\pm}^{z}=\frac{1}{2}\mathrm{Re}\left\{ \int_{0}^{\infty}d\tau e^{\pm i\omega\tau}\mathrm{Tr}_{B}\left\{ \rho_{B}\hat{\Sigma}_{z}\left(\tau\right)\hat{\Sigma}_{z}\left(0\right)\right\} \right\} .\label{eq:gammPmz}
\end{equation}

With this treatment, subtle changes in the bath spectral density from
a frequency shift of order $\sim\omega$ may be taken into account,
without any restrictions on the size of the decay rates themselves.
As we subsequently show, in the EPR-STM experiments discussed in this
work this fact is crucial to interpreting the measured results. The
absence of perturbative assumptions regarding the Rabi frequency gives
rise to an imbalance in the excitation and relaxation rates in the
rotating frame {[}see Eq. \eqref{eq:D0}{]}, translating to an asymmetry
in the EPR lineshape, which was previously not accounted for.

The generalized approach thus results in a more complicated master
equation (see Appendix \ref{4genME} for the full derivation), where
\eqref{eq:BlochN} remains unchanged but \eqref{eq:BlochA} is modified
to 
\begin{align}
\frac{d}{dt}\alpha & =\left[\frac{d}{dt}\alpha\right]_{{\rm Bloch}}-\frac{\Gamma_{+}^{z}-\Gamma_{-}^{z}}{2}\cos\beta\nonumber \\
 & +\left(\Gamma_{+}^{z}+\Gamma_{-}^{z}-2\Gamma^{z}\right)\left[\left(n-\frac{1}{2}\right)\frac{\sin2\beta}{2}-\alpha\cos^{2}\beta\right],\label{eq:generalA}
\end{align}
with $\tan\beta\equiv\frac{\delta\omega}{\Omega}$. It is sufficient
to expand the $\pm\omega$ spectral component around the dc contribution
as $\Gamma_{\pm}^{z}\approx\Gamma^{z}\pm\frac{\Delta}{2}\omega$.
In equilibrium, detailed balance for $\Gamma_{+}^{z}/\Gamma_{-}^{z}$
yields then $\Delta\left(V=0\right)=\frac{\Gamma_{z}}{T}\ll1$. As
we show below, despite the smallness of this correction, its relative
importance is strongly enhanced in the presence of a finite dc bias
$V$, which takes the system out of equilibrium, and may suppress
the even term in the lineshape. The physical origin of $\Delta$ is
in the interference of the electronic continuum in the leads with
the precession of the adatom qubit at frequency $\omega$, as is evident
by the full rate calculations in Appendix \ref{ratesApp}. Plugging
this expansion into the generalized master equation, we find the modified
steady-state polarization, 
\begin{align}
\left\langle \sigma_{z}^{\infty}\right\rangle  & =\ensuremath{\left\langle \sigma_{z}^{\infty}\right\rangle {}_{\mathrm{Bloch}}}+\Delta\frac{\Omega^{2}\delta\omega}{\left(\Gamma_{\downarrow}+\Gamma_{\uparrow}\right)\left(\tilde{\Gamma}^{2}+\delta\omega^{2}\right)+\tilde{\Gamma}\Omega^{2}},\label{eq:GenralSigmaz}
\end{align}
which, compared to \eqref{eq:labframSz}, has an additional contribution,
\textit{odd} in $\delta\omega$. Eq. \eqref{eq:GenralSigmaz}, however,
does not guarantee that the lineshape becomes visibly asymmetric.
Defining $\Delta\sigma\equiv\left\langle \sigma_{z}^{\infty}\right\rangle -\left\langle \sigma_{z}^{\infty}\right\rangle _{\Omega=0}$,
it may be written in a form compliant with Ref. \cite{Willke2018},
Eq. (S2), 
\begin{equation}
\Delta\sigma=\Delta\sigma_{peak}\frac{1+2q^{*}\frac{\delta\omega}{\Gamma_{\Omega}}}{1+\left(\frac{\delta\omega}{\Gamma_{\Omega}}\right)^{2}},
\end{equation}
with $\Delta\sigma_{peak}\equiv-\frac{\Gamma_{\uparrow}-\Gamma_{\downarrow}}{\Gamma_{\downarrow}+\Gamma_{\uparrow}}\frac{\Omega^{2}}{\tilde{\Gamma}\left(\Gamma_{\downarrow}+\Gamma_{\uparrow}\right)+\Omega^{2}}$,
$\Gamma_{\Omega}^{2}\equiv\tilde{\Gamma}^{2}+\frac{\tilde{\Gamma}}{\Gamma_{\downarrow}+\Gamma_{\uparrow}}\Omega^{2}$,
and 
\begin{equation}
q^{*}=\frac{\Delta}{2}\frac{\Gamma_{\downarrow}+\Gamma_{\uparrow}}{\Gamma_{\downarrow}-\Gamma_{\uparrow}}\sqrt{1+\frac{\Omega^{2}}{\tilde{\Gamma}\left(\Gamma_{\downarrow}+\Gamma_{\uparrow}\right)}}.\label{eq:FanoFactor}
\end{equation}
This so-called Fano parameter is a measure of the visibility of asymmetry
in the lineshape. Note that $q^{*}$ grows with $\Omega$, in a manner
consistent with Ref. \cite{Willke2018}, Fig. (S2). Since the square
root term is of the order of $1$, and because $\frac{\Gamma^{z}}{T}\approx10^{-3}$
in this experiment, visible asymmetry requires $\left|\frac{\Gamma_{\downarrow}-\Gamma_{\uparrow}}{\Gamma_{\downarrow}+\Gamma_{\uparrow}}\right|\ll1$,
or $\Gamma_{\downarrow}\approx\Gamma_{\uparrow}$. This is quite unusual
that the excitation and relaxation rates are almost identical, since
$T\ll\omega_{d}$. This points at the crucial role played by the dc
voltage $V$, which is the most dominant energy scale in the system.
Such a scenario is also consistent with the experimental observation
that the measured $T_{1}$ decay time dramatically increases when
this voltage is turned off \cite{STMexperiment}. We will now show
that finite $V$ can indeed make $\Gamma_{\downarrow}\approx\Gamma_{\uparrow}$
(and thus to make $q^{*}$ significant although $\Delta$ is small),
provided one also keeps in mind the distinction between the physical
spin of the adatom and the effective TLS. The latter will also allow
us to explain why $T_{1}\gg T_{2}$.

The different dissipative rates can be calculated explicitly in terms
of the properties of the electronic reservoirs and their couplings
to the TLS. The detailed calculation is straightforward yet lengthy,
and is presented in Appendix \ref{ratesApp}. Under the conditions
$\psi\approx\frac{\pi}{2},$ $\omega\ll T$, $\omega_{d}\ll V,$ and
full polarization of the tip, we obtain \begin{subequations} 
\begin{equation}
\Gamma_{\uparrow}=\frac{\tilde{J_{s}}^{2}}{4}\alpha_{\perp}^{2}T\left[\left(2\chi^{2}+1\right)\left(\eta_{+}\left(\frac{\omega_{d}}{T}\right)+r\right)+2s_{dc}\chi r\right]\label{eq:kFinal}
\end{equation}
\begin{equation}
\Gamma_{\downarrow}=\frac{\tilde{J_{s}}^{2}}{4}\alpha_{\perp}^{2}T\left[\left(2\chi^{2}+1\right)\left(\eta_{-}\left(\frac{\omega_{d}}{T}\right)+r\right)-2s_{dc}\chi r\right]\label{eq:kFinal-1}
\end{equation}
\begin{align}
\Gamma^{z} & =\frac{\tilde{J_{s}}^{2}}{2}\alpha_{zz}^{2}T\left(1+r\right),\label{eq:DeltaInd}
\end{align}
\begin{align}
\Gamma_{\pm}^{z} & =\frac{\tilde{J_{s}}^{2}}{2}\alpha_{zz}^{2}\left[T\left(1+r\right)\pm\omega\right],\label{eq:DeltaInd-1}
\end{align}
\end{subequations} with $\eta_{\pm}\left(x\right)\equiv\pm\frac{x}{e^{\pm x}-1}$,
$\tilde{J_{s}}\equiv\nu_{s}J_{s}$, and $s_{dc}=\pm1$ corresponds
to the dc voltage sign. Note that we find $\Delta=\tilde{J_{s}}^{2}\alpha_{zz}^{2}$.
The parameters $\chi\equiv\frac{\alpha_{zx}}{\alpha_{\perp}}$ and
$r\equiv\frac{\nu_{t}J_{t}}{\nu_{s}J_{s}}\frac{\left|V\right|}{T}$
quantify, respectively, the projection of the adatom Hamiltonian into
the effective TLS and the ratio of the atom interaction strength to
the tip and the substrate. Since $\alpha_{zz}\gg\alpha_{\perp}$,
the relaxation time $T_{1}$ becomes much longer than the dephasing
time $T_{2}$, as found in the experiment. This conclusion originates
in the microscopic treatment of the system, regardless of our modification
of the master equation itself. Note that although our microscopic
consideration produce a ratio $\frac{\alpha_{zz}}{\alpha_{\perp}}\sim4\cdot10^{3}$,
experimental observation of the ratio between relaxation times suggest
that this ratio, while exceptionally large, is realistically about
$1$--$2$ orders of magnitude smaller.

\begin{figure}
\begin{centering}
\includegraphics[bb=0bp 0bp 386bp 264bp,scale=0.55]{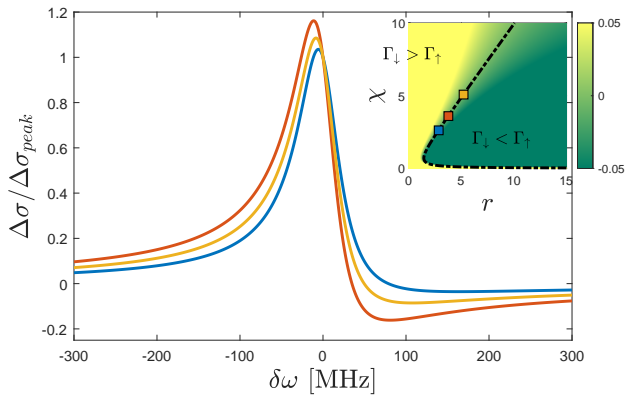} 
\par\end{centering}
\caption{\label{fig:line_shapes} Lineshapes calculated using the proposed
generalized approach with $r=2.83$, $\chi=2.62$ (blue line), $r=3.75$,
$\chi=3.6$ (red line), and $r=5.2$, $\chi=5.08$ (in yellow). Other
parameters: $\Omega=0.8\,\mathrm{MHz}$, $\Gamma^{z}=6\,\mathrm{MHz}$,
$\Gamma_{\downarrow}+\Gamma_{\uparrow}=10\,\mathrm{KHz}$, $s_{dc}=1$,
$\omega_{d}=25\,\mathrm{GHz}$, and $T=12.5\,\mathrm{GHz}\frac{\hbar}{k_{B}}$.
Inset: $\frac{\Gamma_{\downarrow}-\Gamma_{\uparrow}}{\Gamma_{\downarrow}+\Gamma_{\uparrow}}$
as a function of $r$ and $\chi$, Eq. \eqref{eq:kFinal}. At the
dashed lines $\Gamma_{\downarrow}=\Gamma_{\uparrow}$. The colored
markers correspond to the lineshapes.}
\end{figure}

Note that the value of $r$ corresponds to the relative importance
of non-equilibrium (finite bias) enhanced relaxation as compared with
the thermal ones.\textcolor{red}{{} }As for $\chi$, it is quite
sensitive to the exact values of the ligand field Hamiltonian parameters
in Table \ref{tab:ligandParams}, parameters that yield $\chi\approx3$.
Thus, we allow deviations from $\chi\approx3$ so as to approach $\Gamma_{\downarrow}\approx\Gamma_{\uparrow}$.
To explore these relations, in the inset to Fig. \ref{fig:line_shapes}
the $r$ and $\chi$ dependence of the ratio $\left|\frac{\Gamma_{\downarrow}-\Gamma_{\uparrow}}{\Gamma_{\downarrow}+\Gamma_{\uparrow}}\right|$
is plotted. It shows that small values of this ratio are plausible
in a substantial regime near the black dashed line where $\Gamma_{\uparrow}=\Gamma_{\downarrow}$.
Finally, let us note that not only is it required that $\Gamma_{\downarrow}\approx\Gamma_{\uparrow}$,
but also that $\Gamma_{\uparrow}$ be slightly larger than $\Gamma_{\downarrow}$
in order to reproduce the correct lineshape, in the same orientation
of the asymmetry observed in \cite{STMexperiment}.

We may now put everything together, and reproduce the asymmetric lineshapes
using a sensible choice of the different parameters, along with staying
consistent with quantities which were already measured, i.e., the
rough estimates for the decay times and driving intensity that appear
in \cite{STMexperiment}. Fig. \ref{fig:line_shapes} features some
examples of lineshapes that have the same form as in \cite{STMexperiment},
with $r$ and $\chi$ taken such that the asymmetry is visible. The
parameter $q^{*}$ (Eq. \eqref{eq:FanoFactor}) assumes the values
0.2--0.4 for these curves.

\begin{figure}
\begin{centering}
\includegraphics[scale=0.2]{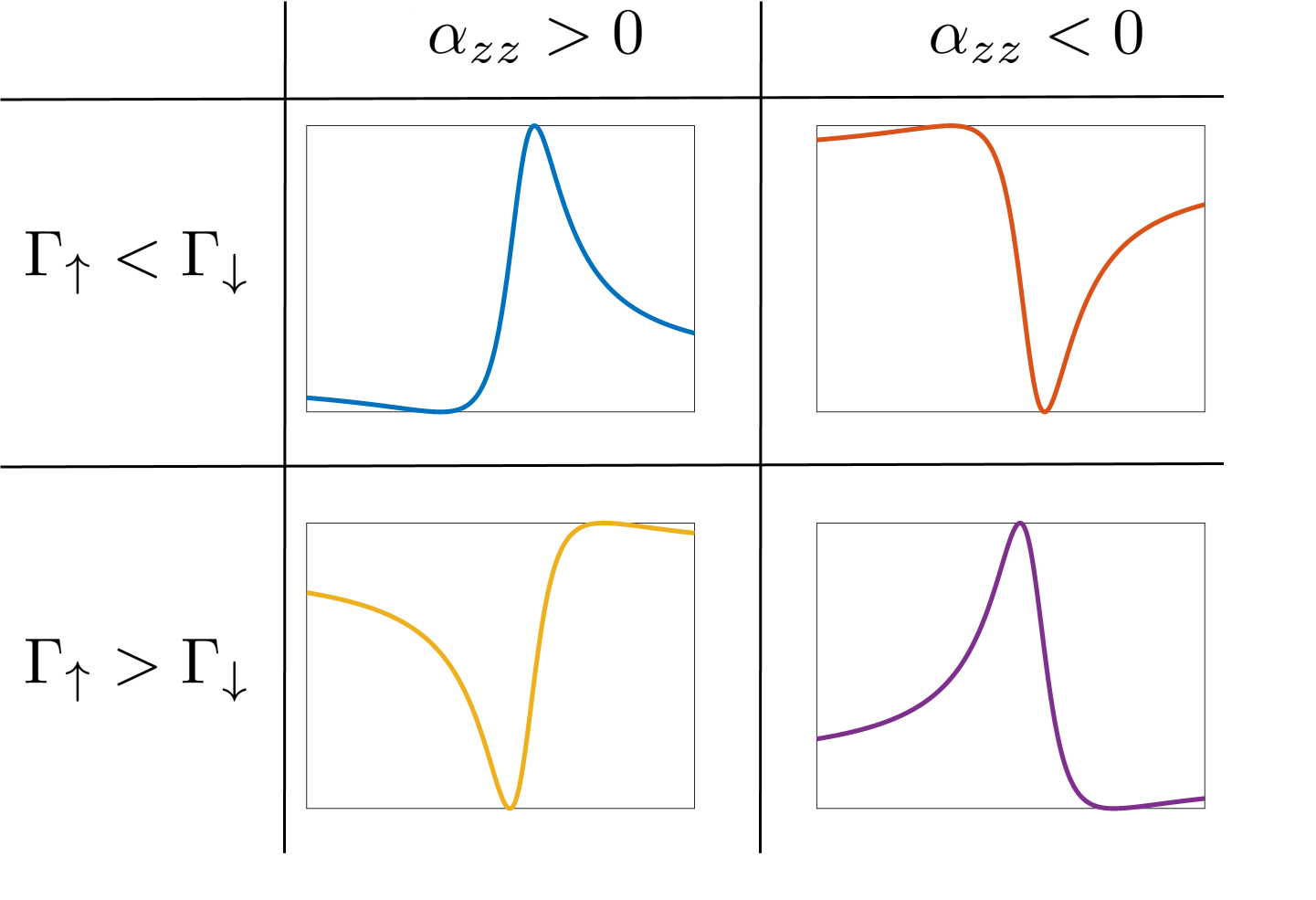} 
\par\end{centering}
\caption{\label{fig:fout_lineshapes} Schematic line shapes given by our generalized
master equation. Depending on the sign of $\left(\Gamma_{\uparrow}-\Gamma_{\downarrow}\right)$
and that of $\alpha_{zz}$, the line shape can take the form of a
positive or negative dip in the tunneling measurement, with an asymmetry
that is ``skewed'' to the right or left of the resonance peak. The
direct dependence of the tunneling current on the sign of $V$ is
neglected to conform with experimental conventions.}
\end{figure}

We note that by changing the parameters the lineshape can be flipped
along either the horizontal or vertical axis. Since $\left\langle I_{T_{0}J}\right\rangle \propto V\alpha_{zz}\left\langle \sigma_{z}\right\rangle $,
changing the sign of $\alpha_{zz}$ due to, e.g., a change in the
direction of $B_{z}$ will lead to a resonant dip instead of a peak,
as was observed in Refs. \cite{negdip,Choi2017}. Moreover, different
microscopical parameters in the system would affect the rates $\Gamma_{\uparrow/\downarrow}$
and could change the sign of $q^{*}$, cf. Ref. \cite{WillkeNew}.
A summary of the possible line shapes is given in Fig. \ref{fig:fout_lineshapes}
below. Notice that the plots in Fig. \ref{fig:line_shapes} correspond
to the lower right quadrangle of Fig. \ref{fig:fout_lineshapes}.

\section{High dc voltage regime}

\label{HighV}

A more comprehensive study of the EPR-STM properties of the system
under discussion was performed in \cite{Willke2018}. Importantly,
unlike the scenario in \cite{STMexperiment} we have discussed thus
far, the energy scale of the bias voltage used was much higher than
the energy separation between the bottom two levels of the adatom
and the higher energy manifold. This makes higher adatom levels accessible,
and seems to complicate our two-level treatment. However, as we will
show below, the relaxation rate to excitation rate ratio for these
levels is large, so their average population is small. This allows
us to perturbatively eliminate them, while renormalizing the rate
constants of the TLS.

\begin{figure}
\begin{centering}
\includegraphics[scale=0.3]{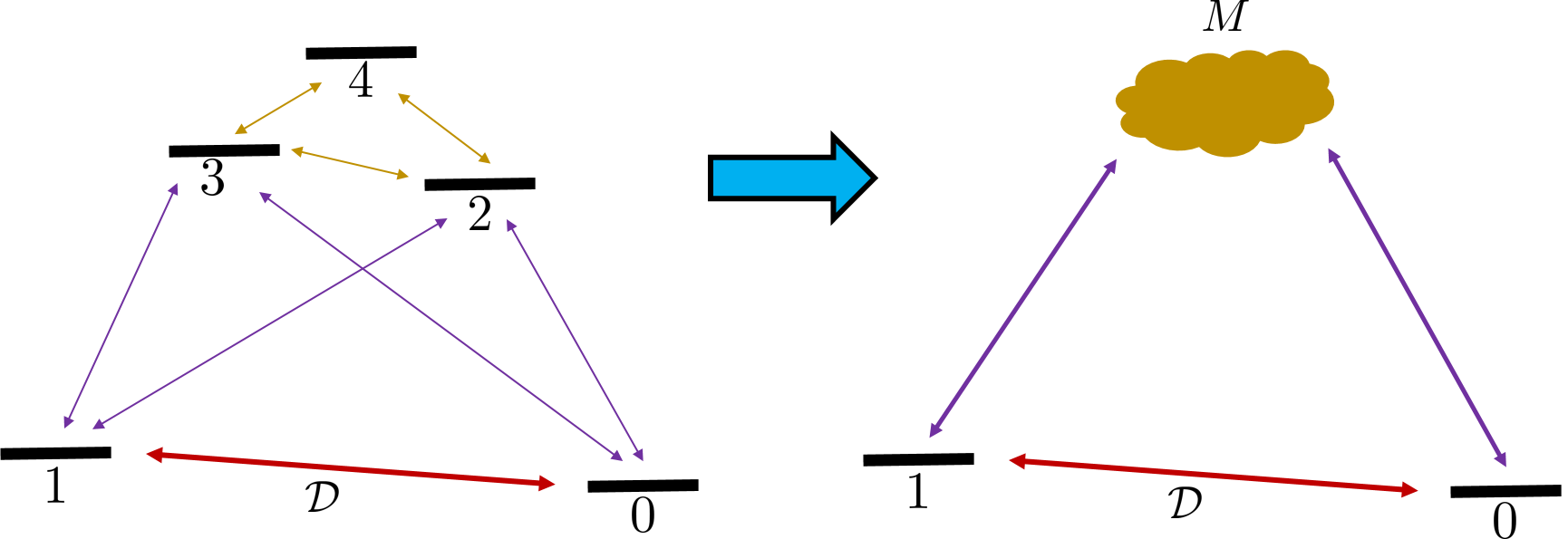} 
\par\end{centering}
\caption{\label{fig:lvl2M}Left: energy levels diagram of the Fe atom, with
the different significant transition processes marked by arrows. The
horizontal ``coordinate'' represents the $\left\langle S_{z}\right\rangle $
of the appropriate level. Right: the simplified diagram used in the
effective master equation, exploiting the spin structure of the energy
level diagram.}
\end{figure}

We employ a simplified scheme, where only consider the transition
rates between different energy levels (see Fig. \ref{fig:lvl2M}),
neglecting effects caused by coherences, which should decay to zero
(except for the coherence of the bottom two levels, already taken
into account). Next, by using the matrix elements $\left\langle i\left|S_{\pm}\right|j\right\rangle $
between each two levels labeled $i,j$, we find the dominant transition
processes. We arrive at the following conclusions:
\begin{itemize}
\item Each of the bottom $|0\rangle,|1\rangle$ levels is strongly connected
to each of the upper $|2\rangle,|3\rangle$ levels, as seen by 
\[
S_{-,02}\approx S_{+,13}\approx0.78,\,\,\,\,-S_{-,03}\approx S_{+,12}\approx0.6.
\]
The magnitudes of the matrix elements with $|4\rangle$ are $\lesssim0.08$,
one order of magnitude weaker, and therefore negligible.
\item The upper levels $|2\rangle,|3\rangle,|4\rangle,$ are all interconnected,
and the transitions $|2\rangle\leftrightarrow|4\rangle$ and $|3\rangle\leftrightarrow|4\rangle$
are of similar amplitude. We find the dominant matrix elements are
$S_{-,23}\approx-0.43,$ $S_{x,24}\approx1.55,$ and $iS_{y,34}\approx1.64.$
\end{itemize}
In light of these observations, we simply approximate the upper levels
as a composite state $M$, with new transition rates, $\Gamma_{0/1\rightarrow M}=\Gamma_{0/1\rightarrow2}+\Gamma_{0/1\rightarrow3}$,
see Fig. \ref{fig:lvl2M}. The master equation for the occupation
properties of level $i$, $P_{i}$, is written as\begin{subequations}
\begin{equation}
\frac{d}{dt}P_{0}=\mathcal{D}_{0}-P_{0}\Gamma_{0\rightarrow M}+P_{M}\Gamma_{M\rightarrow0},\label{eq:dp0-1}
\end{equation}
\begin{equation}
\frac{d}{dt}P_{1}=\mathcal{D}_{1}-P_{1}\Gamma_{1\rightarrow M}+P_{M}\Gamma_{M\rightarrow1},\label{eq:dp1-1}
\end{equation}
\begin{equation}
\frac{d}{dt}P_{M}=-P_{M}\left(\Gamma_{M\rightarrow0}+\Gamma_{M\rightarrow1}\right)+P_{0}\Gamma_{0\rightarrow M}+P_{1}\Gamma_{1\rightarrow M},\label{eq:dp2-1}
\end{equation}
\end{subequations} with $\mathcal{D}_{0}/\mathcal{D}_{1}$ the part
coming from our novel generalized master equation for the TLS, Eqs.
\eqref{eq:BlochN},\eqref{eq:BlochA},\eqref{eq:generalA}, and $\Gamma_{i\rightarrow j}$
the transition rate form level $i$ to level $j$. In the steady state,
$\frac{d}{dt}P_{M}=0$, we find the inclusion of the $M$ composite
state results in a modification of the relaxation and excitation rates
for the TLS, 
\begin{equation}
\Gamma_{\downarrow}\rightarrow\Gamma_{\downarrow}+\frac{\Gamma_{1\rightarrow M}\Gamma_{M\rightarrow0}}{\Gamma_{M\rightarrow0}+\Gamma_{M\rightarrow1}},\label{eq:gammaM1}
\end{equation}
\begin{equation}
\Gamma_{\uparrow}\rightarrow\Gamma_{\uparrow}+\frac{\Gamma_{0\rightarrow M}\Gamma_{M\rightarrow1}}{\Gamma_{M\rightarrow0}+\Gamma_{M\rightarrow1}}.\label{eq:gammaM2}
\end{equation}

The transition rates can then be evaluated in a similar manner to
the scheme used in Appendix \ref{ratesApp}. For each $\Gamma_{i\rightarrow j}$
we calculate the $\bar{\alpha}$ matrix that connects the physical
spin operator to a two-level representation of levels $i,j,$ expressed
as $\vec{S}=\bar{\alpha}_{\left\langle i,j\right\rangle }\vec{\sigma}$.
The Pauli matrices $\vec{\sigma}$ represent the reduced Hilbert space
of the two levels $\left(i,j\right)$. The energy difference between
each two relevant levels $\Delta E_{i,j}$ is take into account, and
thermal contributions to excitation rates (i.e., ones that do not
involve the bias voltage) are neglected, since $\Delta E\gg T$ when
any of the higher energy levels are involved. With $\psi\approx81^{\circ}$
and $V=60$ mV, which are the parameters values in the relevant high
voltage experiment \cite{Willke2018}, we find 
\begin{align*}
\Gamma_{M\rightarrow0} & \approx0.73\bar{\Gamma}\left(1+6r_{ts}+r_{ts}^{2}\right),\,\,\,\Gamma_{0\rightarrow M}\approx1.67\bar{\Gamma}r_{ts},
\end{align*}
\begin{align*}
\Gamma_{M\rightarrow1} & \approx0.73\bar{\Gamma}\left(1+4.7r_{ts}+r_{ts}^{2}\right),\,\,\,\Gamma_{1\rightarrow M}\approx2.34\bar{\Gamma}r_{ts},
\end{align*}
where $r_{ts}\equiv\frac{J_{t}\nu_{t}}{J_{s}\nu_{s}}$, $\bar{\Gamma}\equiv2.15\frac{\alpha_{\perp}^{2}}{4}J_{s}^{2}\nu_{s}^{2}\Delta E$.
We observe that the ratio between the rate of exciting the adatom
into the higher energy manifold, compared to the relaxation rate out
of it, is of the order $r_{ts}=r\frac{T}{V}$. Since $\frac{T}{V}\sim10^{-3}$,
and we find that reproducing the experimental results dictates $r\sim10$,
our assumption of very low occupation for the high energy levels is
well-justified. Crucially, we find the rates $\Gamma_{0\rightarrow2}$
and $\Gamma_{1\rightarrow3}$ are of comparable size, contradicting
the existence of an appreciable spin-torque effect, where transitions
between levels lowering $\left\langle S_{z}\right\rangle $ are favored
compared to ones raising it, or vice versa, depending on the sign
of the dc voltage. The comparable size of theses transition rates
can be traced to the fact that whereas in our earlier analysis for
the $|0\rangle,|1\rangle$ levels, $\frac{\alpha_{zx}}{\alpha_{\perp}}\equiv\chi$
was an $\mathcal{O}\left(1\right)$ number (around 3), for the upper
levels it is $\mathcal{O}\left(10^{-2}\right).$ This, in conjunction
with $\cos\psi$ being close to zero, strongly attenuates the $\left\langle S_{z}\right\rangle $
directionality of the inter-level transitions.

\begin{figure}
\begin{centering}
\includegraphics[bb=50bp 0bp 570bp 550bp,scale=0.48]{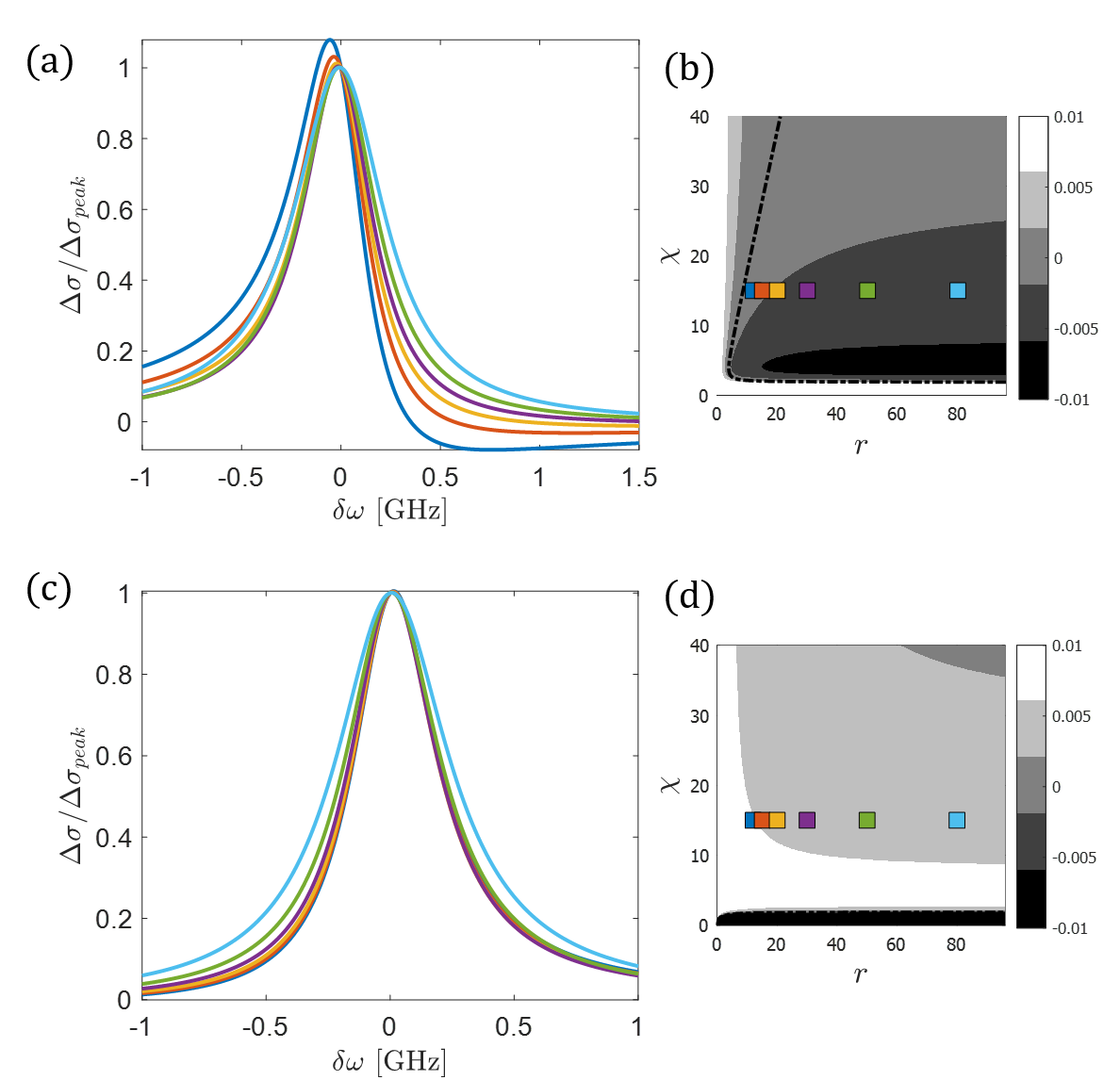}
\par\end{centering}
\caption{\label{fig:HighVline} (a) Lineshapes in the high voltage regime for
positive bias voltage, with different values of $r$, representing
different tip-adatom separations, for $\chi=15$. (b) $\frac{\Gamma_{\downarrow}-\Gamma_{\uparrow}}{\Gamma_{\downarrow}+\Gamma_{\uparrow}}$
with a positive voltage, as a function of $r$ and $\chi$ with colored
markers corresponding to the different lineshapes in (a). (c), (d)
are the same as (a), (b), respectively, for negative bias voltage.
Other parameters (based on \cite{Willke2018}): $\Omega=20\,\mathrm{MHz}$,
$\Gamma^{z}=25\,\mathrm{MHz}$, $\Gamma_{\downarrow}+\Gamma_{\uparrow}=0.5\,\mathrm{MHz}$,
$\bar{a}=15,$ $\gamma=0.25$, $\omega_{0}=21.5\,\mathrm{GHz}$, and
$T=25\,\mathrm{GHz}\frac{\hbar}{k_{B}}$. Case (a) corresponds tot
the lower right quadrangle of Fig. \ref{fig:fout_lineshapes}, while
case (c) to the upper right one (note the renormalization by $\Delta\sigma_{peak}$
flips the lineshape sign).}
\end{figure}

The effective TLS master equation now has, up to second order corrections
in the small parameter $r_{ts}$, 
\begin{equation}
\Gamma_{\downarrow}\rightarrow\Gamma_{\downarrow}+\bar{\Gamma}\left(1+\gamma\right),
\end{equation}
\begin{equation}
\Gamma_{\uparrow}\rightarrow\Gamma_{\uparrow}+\bar{\Gamma}\left(1-\gamma\right),
\end{equation}
where in our calculations $\gamma\approx0.16$. Thus, by accounting
for the microscopic details of the adatom spin matrix elements, we
see that both the TLS excitation and relaxation rates are increased
with \textit{comparable magnitude}. We note that reversing the bias
direction will amount to taking $\gamma\rightarrow-\gamma$ in the
above modification. We finally write the high voltage regime relaxation
and excitation rates as 
\begin{align}
\Gamma_{\uparrow/\downarrow} & =\frac{\tilde{J_{s}}^{2}}{4}\alpha_{\perp}^{2}T\left(2\chi^{2}+1\right)\left(\eta_{\pm}\left(\frac{\omega_{d}}{T}\right)+r\left(1+\bar{a}\right)\right)\nonumber \\
 & \mp s_{dc}\frac{\tilde{J_{s}}^{2}}{4}\alpha_{\perp}^{2}T\left(\bar{a}\gamma-2\chi\right)r\label{eq:kHighV}
\end{align}
where $\bar{a}$ is the relative amplitude of the $M$-assisted transitions
compared to the direct ones $\bar{a}\approx\frac{2\bar{\Gamma}}{\Gamma_{\downarrow}^{ts}+\Gamma_{\uparrow}^{ts}}$,
with $\Gamma^{ts}$ being the tip-atom-substrate tunneling contribution.
We note that by taking $\bar{a}=0$, one recovers the rates of the
low voltage regime, Eqs. \eqref{eq:kFinal}--\eqref{eq:kFinal-1}.
The critical line where $\Gamma_{\uparrow}=\Gamma_{\downarrow}$ depends
on the voltage sign and is given by
\begin{equation}
r=s_{dc}\frac{\omega_{d}}{2T}\frac{2\chi^{2}+1}{2\chi-\bar{a}\gamma}.\label{eq:criticallINES}
\end{equation}
To estimate the size of $\bar{a}$, we examine for example 
\[
\frac{\bar{\Gamma}}{\Gamma_{\downarrow}^{ts}}\approx\frac{\alpha_{\perp\left\langle 0,2\right\rangle }^{2}}{\alpha_{\perp\left\langle 0,1\right\rangle }^{2}}\times\frac{1}{\left(2\chi^{2}-2\chi+1\right)2}.
\]
The righthand fraction has the order of $10^{-2}$, as $\chi\sim3-8$
in our analysis thus far. However the lefthand fraction seems huge
and of order $10^{6}$. Actually we know that $\alpha_{\perp\left\langle 0,1\right\rangle }^{2}$
is much larger than the order $10^{-3}$ evaluated, as discussed above
in Sec. \ref{generalizedApproach}. Taking this into account, we may
estimate that $\bar{\Gamma}$ is of the same order as $\Gamma_{\downarrow}$
and $\Gamma_{\uparrow}$ or perhaps one order of magnitude larger,
leading to an estimate of $\bar{a}\sim10$.

\begin{figure}
\begin{centering}
\includegraphics[bb=0bp 0bp 650bp 220bp,scale=0.45]{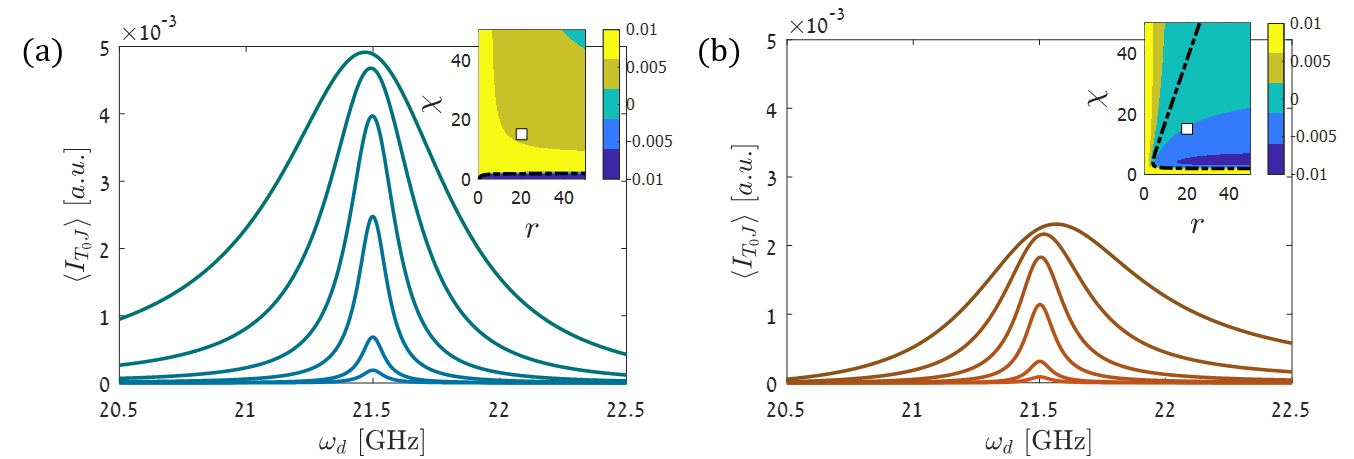} 
\par\end{centering}
\caption{\label{fig:HighVlineS} Tunneling line shapes in the high voltage
regime for different signs dc bias voltage {[}(a) positive, (b) negative{]},
with increasing driving amplitude (bottom to top) = $1,2,5,10,20,40$
MHz. We use $\chi=15,$ $r=20$ in both cases. Other parameters (based
on the experimental results in \cite{Willke2018}): $\Gamma^{z}=25\,\mathrm{MHz}$,
$\Gamma_{\downarrow}+\Gamma_{\uparrow}=0.5\,\mathrm{MHz}$, $\bar{a}=15,$
$\gamma=0.25$, $\omega_{d}=21.5\,\mathrm{GHz}$, and $T=25\,\mathrm{GHz}\frac{\hbar}{k_{B}}$.
Insets: $\frac{\Gamma_{\downarrow}-\Gamma_{\uparrow}}{\Gamma_{\downarrow}+\Gamma_{\uparrow}}$
for different signs of the voltage, as a function of $r$ and $\chi$
with the working point $(20,15)$ marked by a square.}
\end{figure}

The form of $\Gamma_{\uparrow/\downarrow}$ we find in Eq. \eqref{eq:kHighV}
allows us to reproduce the main features of the experiment in Ref.
\cite{Willke2018}. As an example, the observed change in the direction
of the lineshape asymmetry with reversal of the bias voltage (see
Supplementary Material for Ref. \cite{Willke2018}) is recreated in
Fig. \ref{fig:HighVline}, with $\left|q^{*}\right|$ values as high
as $\sim0.3$. Changing the voltage subsequently affects $\Gamma_{\uparrow}/\Gamma_{\downarrow},$
enabling a scenario where one flips the sign $\left(\Gamma_{\uparrow}-\Gamma_{\downarrow}\right)$,
and subsequently that of $q^{*}$, determining the asymmetry direction.
Note that the very different critical dashed lines in the two insets
of Fig. \ref{fig:HighVline} are given by Eq. \eqref{eq:criticallINES}.
Moreover, the widening of the resonance with a decrease in the tip-atom
separation is also apparent. This is encoded by an increase in $r$,
which is proportional to the amplitude $J_{t}$. The decoherence rate
$\Gamma^{z}$ increases with $r$, naturally leading to a wider lineshape.
Additionally, the reported rise in asymmetry as the driving amplitude
is increased is reproduced {[}Eq. \eqref{eq:FanoFactor}{]}. This
is evident in Fig. \ref{fig:HighVlineS}, where the line shapes are
calculated with varying driving amplitude. One should compare this
with Figs. S2 and S5B in the Supplementary Material for \cite{Willke2018},
which clearly show similar features.

\section{Conclusions}

\label{cocncluze}

In conclusion, we have shown that treating the adatoms in spin-polarized
STM-EPR experiments as a driven open quantum system requires special
care and a novel generalized approach. On the one hand, this approach
should not treat the driving in a perturbative manner, such that changes
in the bath spectral density as a result of small frequency shifts
are resolved by the different master equation rates. On the other
hand, it should be valid in the experimental parameter regime, where
some of the decay rates may be significantly large compared to the
driving amplitude.

We present such a treatment, which allows a clear understanding of
the origin of the asymmetry in recorded lineshapes, due to the small
difference in the electronic reservoirs spectral functions at $0$
and $\pm\omega$. We find that even a modest difference allows for
a small odd component in the lineshape {[}Eq. \eqref{eq:GenralSigmaz}{]},
whose relative importance is greatly enhanced by the dc bias in the
experiments considered.

As we have shown, completely accounting for the underlying physics
of the experimental system is crucial: We find that tuning the voltage
as to make the TLS relaxation and excitation rates close strongly
suppresses the even component of the lineshape {[}Eq. \eqref{eq:labframSz}{]},
leading to the observed asymmetry. Moreover, projecting the physical
spin onto the TLS description allows one to understand the origin
of the large $T_{1}/T_{2}$ ratio, namely the dominance of the matrix
element $\alpha_{zz}$, relating spin projection along the lab $\hat{z}$
direction to the TLS polarization.

Furthermore, our novel approach enables recreating virtually all other
experimental trends, even in higher voltage regimes, e.g., a change
of the asymmetry sign depending on the bias voltage or on the magnetic
field orientation \cite{Willke2018,Choi2017,CHOI_ESR_STM,WillkeNew},
and the dependence of the line shapes asymmetry on driving amplitude.

The generalized approach developed here may be useful in properly
analyzing results from any future EPR-STM studies, as well as for
other open quantum systems which involve non-trivial parametric regimes,
e.g., hybrid quantum devices \cite{HybridSystemsKlaus,Hybrid1,Hybrid2}.
\begin{acknowledgments}
M. G. was supported by the Israel Science Foundation (Grant No. 227/15),
the German Israeli Foundation (Grant No. I-1259-303.10), the US-Israel
Binational Science Foundation (Grants No. 2014262 and 2016224), and
the Israel Ministry of Science and Technology (Contract No. 3-12419).
B. H. acknowledges support by German-Israeli DIP project (Hybrid devices:
FO 703/2--1).
\end{acknowledgments}

\appendix

\section{Derivation of the generalized master equation}

\label{4genME}

We derive our generalized master equation so as to solve the non-secular
problem with respect to the frequencies $0,\omega$, a situation that
is essential for the experimental case with $\frac{1}{T_{1}}<\Omega<\frac{1}{T_{2}}$.
The starting Hamiltonian has the form (setting $\hbar=1$) 
\begin{align}
H & =-\frac{1}{2}\omega_{0}\sigma_{z}+\frac{\Omega}{2}\left(e^{i\omega_{d}t}\sigma_{+}+e^{-i\omega_{d}t}\sigma_{-}\right),\nonumber \\
 & -\frac{1}{2}\left(a_{x}\sigma_{x}+a_{y}\sigma_{y}+a_{z}\sigma_{z}\right)\otimes\hat{B}+H_{B}\label{eq:Bhamilton}
\end{align}

\noindent with $\omega_{0}$ the two-level energy separation, $\Omega$
the driving intensity, $\omega_{d}$ the driving frequency (and $\delta\omega\equiv\omega_{d}-\omega_{0}$),
and $H_{B}$ the bath Hamiltonian. For simplicity we will first consider
the case where all the impurity operators couple to the same bath
operator $\hat{B}$, and later on extend our results to the more general
case of different bath operators. The coefficients $a_{x,y,z}$ represent
some general form of coupling to the bath; we also define $a\equiv a_{x}+ia_{y}$
and $a_{0}\equiv a_{z}$ . We apply a transformation to a rotating
frame by defining $U\equiv e^{-\frac{i\omega_{d}t}{2}\sigma_{z}}$,
\begin{equation}
H\rightarrow UHU^{\dagger}+i\dot{U}U^{\dagger}.
\end{equation}
Our transformed Hamiltonian reads 
\begin{align}
H & =\frac{1}{2}\delta\omega\sigma_{z}+\frac{1}{2}\Omega\sigma_{x}\nonumber \\
 & -\frac{1}{2}\left(ae^{i\omega_{d}t}\sigma_{-}+a^{*}e^{-i\omega_{d}t}\sigma_{+}+a_{0}\sigma_{z}\right)\hat{B}+H_{B}.
\end{align}
We now diagonalize the system Hamiltonian using the transformation
$\tilde{H}=S^{-1}HS$, using 
\begin{equation}
S=\frac{1}{\sqrt{2}}\underbrace{\begin{pmatrix}\cos\frac{\beta}{2}+\sin\frac{\beta}{2} & -\cos\frac{\beta}{2}+\sin\frac{\beta}{2}\\
\cos\frac{\beta}{2}-\sin\frac{\beta}{2} & \cos\frac{\beta}{2}+\sin\frac{\beta}{2}
\end{pmatrix}}_{\mathrm{system\,subspace}}\otimes\mathbb{1}_{\mathrm{bath\,subspace}}\label{eq:SmatrixDiagonal}
\end{equation}
with the angle $\beta$ defined by $\tan\beta\equiv\frac{\delta\omega}{\Omega}$.
Applying this transformation we get 
\begin{equation}
\tilde{H}=\frac{1}{2}\omega\sigma_{z}-\left(A_{0}+A_{1}+A_{-1}\right)\hat{B}+H_{E},\label{eq:DiagonalFrame}
\end{equation}
with\begin{subequations} 
\begin{equation}
A_{0}=\left(\frac{\sin\beta}{2}a_{0}+\frac{\cos\beta}{4}\left(a^{*}e^{-i\omega_{d}t}+ae^{i\omega_{d}t}\right)\right)\sigma_{z},
\end{equation}
\begin{equation}
A_{1}=\left(\frac{\sin\beta-1}{4}a^{*}e^{-i\omega_{d}t}+\frac{\sin\beta+1}{4}ae^{i\omega_{d}t}-\frac{\cos\beta}{2}a_{0}\right)\sigma_{-},
\end{equation}
\begin{equation}
A_{-1}=\left(A_{1}\right)^{\dagger}.
\end{equation}
\end{subequations}

Now that we have obtained a Hamiltonian with a diagonal system term,
we move into the interaction picture, with the additional time dependence
$A_{k}\left(t\right)\rightarrow e^{ik\omega t}A_{k}\left(t\right)$,
where $\omega=\sqrt{\Omega^{2}+\delta\omega^{2}}$ is the generalized
Rabi frequency. We may now use the Markovian expression for the time
evolution of the reduced system density matrix, 
\begin{align}
\frac{d}{dt}\tilde{\rho}\left(t\right) & =\sum_{j,k=-1}^{1}\int_{0}^{\infty}ds\mathrm{Tr}_{B}\left\{ \rho_{B}\hat{B}\left(s\right)\hat{B}\left(0\right)\right\} \times\nonumber \\
 & \left[A_{j}\left(t-s\right)\tilde{\rho}\left(t\right)A_{k}^{\dagger}\left(t\right)-A_{k}^{\dagger}\left(t\right)A_{j}\left(t-s\right)\tilde{\rho}\left(t\right)\right]+\mathrm{h.c.\,,}\label{eq:Markovian}
\end{align}
with $\tilde{\rho}\equiv\frac{d+u}{2}+\frac{d-u}{2}\sigma_{z}+x\sigma_{-}+x^{*}\sigma_{+}$,
the density matrix in the basis of $\tilde{H}$, which is different
than the original (non-diagonal) ``lab frame'' basis. Eq. \eqref{eq:Markovian},
upon neglecting terms oscillating with the high frequencies $\pm\omega_{d},\pm\omega_{d}\pm\omega,\pm2\omega_{d}$
(the usual secular approximation while keeping frequencies $0,\omega,2\omega$),
leads to the master equation 
\begin{align}
\frac{d}{dt}\tilde{\rho}\left(t\right) & =\mathcal{D}_{0}+\mathcal{D_{\omega}}+\mathcal{D}_{2\omega}+\mathrm{h.c.},\label{eq:rhoByDs}
\end{align}
with 
\begin{widetext}
\begin{subequations} 
\begin{eqnarray}
\mathcal{D}_{0} & \equiv & \left(-d\sigma_{z}-x^{*}\sigma_{+}\right)\left|a\right|^{2}\left(\left(\frac{\sin\beta-1}{4}\right)^{2}\Gamma\left(\omega_{d}-\omega\right)+\left(\frac{\sin\beta+1}{4}\right)^{2}\Gamma\left(-\omega_{d}-\omega\right)\right)\nonumber \\
 &  & +\left(-d\sigma_{z}-x^{*}\sigma_{+}\right)\left(\frac{\cos\beta}{2}\right)^{2}a_{0}^{2}\Gamma\left(-\omega\right)\nonumber \\
 &  & +\left(u\sigma_{z}-x\sigma_{-}\right)\left|a\right|^{2}\left(\left(\frac{\sin\beta+1}{4}\right)^{2}\Gamma\left(\omega_{d}+\omega\right)+\left(\frac{\sin\beta-1}{4}\right)^{2}\Gamma\left(-\omega_{d}+\omega\right)\right)\nonumber \\
 &  & +\left(u\sigma_{z}-x\sigma_{-}\right)\left(\frac{\cos\beta}{2}\right)^{2}a_{0}^{2}\Gamma\left(\omega\right)\nonumber \\
 &  & -\left(x\sigma_{-}+x^{*}\sigma_{+}\right)\left(\frac{\sin^{2}\beta}{2}a_{0}^{2}\Gamma\left(0\right)+\left|a\right|^{2}\frac{\cos^{2}\beta}{8}\left(\Gamma\left(\omega_{d}\right)+\Gamma\left(-\omega_{d}\right)\right)\right),\label{eq:D0}
\end{eqnarray}
\begin{align}
D_{\omega}\equiv & e^{i\omega t}\left(-d\sigma_{-}\right)\left|a\right|^{2}\cos\beta\left(\frac{\sin\beta-1}{8}\Gamma\left(\omega_{d}-\omega\right)+\frac{\sin\beta+1}{8}\Gamma\left(-\omega_{d}-\omega\right)\right)\nonumber \\
 & +e^{i\omega t}\left(-d\sigma_{-}\right)\frac{\cos\beta}{2}\left(-\sin\beta a_{0}^{2}\Gamma\left(-\omega\right)\right)\nonumber \\
 & +e^{-i\omega t}u\sigma_{+}\left|a\right|^{2}\cos\beta\left(\frac{\sin\beta+1}{8}\Gamma\left(\omega_{d}+\omega\right)+\frac{\sin\beta-1}{8}\Gamma\left(-\omega_{d}+\omega\right)\right)\nonumber \\
 & +e^{-i\omega t}u\sigma_{+}\frac{\cos\beta}{2}\left(-\sin\beta a_{0}^{2}\Gamma\left(\omega\right)\right)\nonumber \\
 & +e^{-i\omega t}\left[-\sigma_{+}-x\sigma_{z}\right]\frac{\cos\beta}{4}\left(\frac{\sin\beta+1}{4}\left|a\right|^{2}\Gamma\left(-\omega_{d}\right)+\frac{\sin\beta-1}{4}\left|a\right|^{2}\Gamma\left(\omega_{d}\right)-\sin\beta a_{0}^{2}\Gamma\left(0\right)\right)\nonumber \\
 & +e^{i\omega t}\left[\sigma_{-}-x^{*}\sigma_{z}\right]\frac{\cos\beta}{4}\left(\frac{\sin\beta-1}{4}\left|a\right|^{2}\Gamma\left(-\omega_{d}\right)+\frac{\sin\beta+1}{4}\left|a\right|^{2}\Gamma\left(\omega_{d}\right)-\sin\beta a_{0}^{2}\Gamma\left(0\right)\right),
\end{align}
\begin{eqnarray}
\mathcal{D}_{2\omega} & \equiv & e^{2i\omega t}x^{*}\sigma_{-}\left|a\right|^{2}\left(\frac{\sin^{2}\beta-1}{16}\Gamma\left(\omega_{d}-\omega\right)+\frac{\sin^{2}\beta-1}{16}\Gamma\left(-\omega_{d}-\omega\right)\right)\nonumber \\
 &  & +e^{2i\omega t}x^{*}\sigma_{-}\left(\frac{\cos\beta}{2}\right)^{2}a_{0}^{2}\Gamma\left(-\omega\right)\nonumber \\
 &  & +e^{-2i\omega t}x\sigma_{+}\left|a\right|^{2}\left(\frac{\sin^{2}\beta-1}{16}\Gamma\left(\omega_{d}+\omega\right)+\frac{\sin^{2}\beta-1}{16}\Gamma\left(-\omega_{d}+\omega\right)\right)\nonumber \\
 &  & +e^{-2i\omega t}x\sigma_{+}\left(\frac{\cos\beta}{2}\right)^{2}a_{0}^{2}\Gamma\left(\omega\right),
\end{eqnarray}
\end{subequations} 
\end{widetext}

and we defined the bath correlation functions 
\begin{equation}
\Gamma\left(\nu\right)\equiv\mathrm{Re}\left\{ \int_{0}^{\infty}d\tau e^{i\nu\tau}\mathrm{Tr}_{B}\left\{ \rho_{B}\hat{B}\left(\tau\right)\hat{B}\left(0\right)\right\} \right\} .
\end{equation}

Let us note that for vanishing driving, $\Omega=0$, one gets $\omega=\delta\omega$,
$\beta=\pi/2$, hence only the spectral functions at frequencies 0
and $\pm(\omega_{d}+\delta\omega)=\pm\omega_{0}$ remain, so the dependence
on $\omega_{d}$ disappears, as it should. The conventional secular
approximation would now have allowed us to discard $\mathcal{D}_{\omega}$
and $\mathcal{D}_{2\omega}$ terms, but in this generalized treatment
we keep these non-secular terms. In our next step, we perform a unitary
transformation on the master equation \eqref{eq:rhoByDs} with $e^{-i\tilde{H}_{S}t}\left(...\right)e^{i\tilde{H}_{S}t}$
(where $\tilde{H}_{S}=\frac{1}{2}\omega\sigma_{z}$). This introduces
the coherent time evolution term $-i\left[\tilde{H}_{S},\tilde{\rho}\right]$
into the righthand side of \eqref{eq:rhoByDs}, while allowing us
to eliminate the $e^{\pm i\omega t}/e^{\pm2i\omega t}$ time dependence
appearing in $\mathcal{D}_{\omega}/\mathcal{D}_{2\omega}$. The novel
non-secular terms are now more manageable, as they do not introduce
any new time dependencies into the master equation.

For the purposes of this work we may approximate $\Gamma\left(\pm\omega_{d}\pm\omega\right)\approx\Gamma\left(\pm\omega_{d}\right)$,
since $\omega\ll\omega_{d}$, and hence deviations from this assumptions
have only a minor effect on the results we present. We now define
the relevant rates \begin{subequations} 
\begin{equation}
\Gamma_{\downarrow}\equiv\frac{\left|a\right|^{2}}{2}\Gamma\left(\omega_{d}\right),\label{eq:kappaDef}
\end{equation}
\begin{equation}
\Gamma_{\uparrow}\equiv\frac{\left|a\right|^{2}}{2}\Gamma\left(-\omega_{d}\right),\label{eq:kappaeDef}
\end{equation}
\begin{equation}
\Gamma^{z}\equiv\frac{a_{0}^{2}}{2}\Gamma\left(0\right),\label{eq:kappa0Def}
\end{equation}
\begin{equation}
\Gamma_{\pm}^{z}\equiv\frac{a_{0}^{2}}{2}\Gamma\left(\pm\omega\right).\label{eq:kappapmzDef}
\end{equation}
\end{subequations}Plugging these in, and using the inverse of \eqref{eq:SmatrixDiagonal}
to get the master equation in the original basis $\rho\equiv\frac{1}{2}+\left(n-\frac{1}{2}\right)\sigma_{z}+\alpha^{*}\sigma_{-}+\alpha\sigma_{+}$,
one finds that in a frame rotating with frequency $\omega_{d}$ (where
$\alpha$ is transformed as $\alpha e^{-i\omega_{d}t}\rightarrow\alpha$),
the master equation is \begin{subequations} 
\begin{eqnarray}
\frac{d}{dt}n & = & -n\left(\Gamma_{\downarrow}+\Gamma_{\uparrow}\right)+\Gamma_{\uparrow}-i\Omega\frac{\alpha-\alpha^{*}}{2},\label{eq:generalN}
\end{eqnarray}
\begin{eqnarray}
\frac{d}{dt}\alpha & = & -\alpha\left(\frac{\Gamma_{\downarrow}+\Gamma_{\uparrow}}{2}+\left(\Gamma_{+}^{z}+\Gamma_{-}^{z}\right)\cos^{2}\beta+2\Gamma^{z}\sin^{2}\beta+i\delta\omega\right)\nonumber \\
 &  & -i\Omega\left(n-\frac{1}{2}\right)+\left(n-\frac{1}{2}\right)\sin\beta\cos\beta\left(\Gamma_{+}^{z}+\Gamma_{-}^{z}-2\Gamma^{z}\right)\nonumber \\
 &  & -\frac{\Gamma_{+}^{z}-\Gamma_{-}^{z}}{2}\cos\beta.\label{eq:generalAS}
\end{eqnarray}
\end{subequations} By setting $\Gamma_{+}^{z}=\Gamma_{-}^{z}=\Gamma^{z}$
the equations reduce to the standard Bloch equations Eqs. \eqref{eq:BlochN}--\eqref{eq:BlochA}.

Throughout our discussion it was assumed (for reasons of convenience)
that the coupling was via the same bath operator \textbf{$\hat{B}$},
coupled to the system degrees of freedom via general coefficients.
This need not necessarily be the case, as each $\hat{\sigma}$ operator
can generally \textit{couple to a different bath operator}. We could
generalize the Hamiltonian used in Eq. \eqref{eq:Bhamilton} to 
\begin{align}
H & =-\frac{1}{2}\hbar\omega_{0}\sigma_{z}+\frac{\Omega}{2}\left(e^{i\omega_{d}t}\sigma_{+}+e^{-i\omega_{d}t}\sigma_{-}\right)\nonumber \\
 & -\frac{1}{2}\left(\sigma_{+}\hat{\Sigma}_{+}+\sigma_{-}\hat{\Sigma}_{-}+\sigma_{z}\hat{\Sigma}_{z}\right)+H_{B},
\end{align}

\noindent Performing the prescribed diagonalization process, we get

\begin{equation}
\tilde{H}=\frac{1}{2}\omega\sigma_{z}-\left(A_{0}+A_{1}+A_{-1}\right)+H_{B},
\end{equation}

\noindent as before, though with the newly defined\begin{subequations}
\begin{equation}
A_{0}=\left(\frac{\sin\beta}{2}a_{z}\hat{\Sigma}_{z}+\frac{\cos\beta}{4}\left(\hat{\Sigma}_{-}e^{-i\omega_{d}t}+\hat{\Sigma}_{+}e^{i\omega_{d}t}\right)\right)\sigma_{z},
\end{equation}
\begin{equation}
A_{1}=\left(\frac{\sin\beta-1}{4}\hat{\Sigma}_{-}e^{-i\omega_{d}t}+\frac{\sin\beta+1}{4}\hat{\Sigma}_{+}e^{i\omega_{d}t}-\frac{\cos\beta}{2}a_{z}\hat{\Sigma}_{z}\right)\sigma_{-},
\end{equation}
\end{subequations} and $A_{-1}=\left(A_{1}\right)^{\dagger}$. At
first glance this seems to somewhat complicate things: whereas earlier
all the correlation functions we needed to calculate were of the form
$\left\langle \hat{B}\left(\tau\right)\hat{B}\left(0\right)\right\rangle $,
it seems now that correlations such as $\left\langle \hat{\Sigma}_{\pm}\left(\tau\right)\hat{\Sigma}_{z}\left(0\right)\right\rangle $
also need to be taken into account. Luckily, this is not the case.
Due to the lab frame secular approximation, where oscillations by
$\pm\omega_{d}$ (or higher frequency) in time $t$ are neglected,
we are left only with three different bath correlation functions:
$\left\langle \hat{\Sigma}_{-}\left(\tau\right)\hat{\Sigma}_{+}\left(0\right)\right\rangle $,
$\left\langle \hat{\Sigma}_{+}\left(\tau\right)\hat{\Sigma}_{-}\left(0\right)\right\rangle $,
and $\left\langle \hat{\Sigma}_{z}\left(\tau\right)\hat{\Sigma}_{z}\left(0\right)\right\rangle $.
All other mixed products multiply terms which are negligible thanks
to rapid oscillations at higher frequencies. This in turn ensures
that \textit{the structure of the generalized master equation remains
unchanged}, but with more general expressions for the rates, given
by \begin{subequations} 
\begin{equation}
\Gamma_{\downarrow}\equiv\frac{1}{2}\mathrm{Re}\left\{ \int_{0}^{\infty}d\tau e^{i\omega_{d}\tau}\mathrm{Tr}_{B}\left\{ \rho_{B}\hat{\Sigma}_{-}\left(\tau\right)\hat{\Sigma}_{+}\left(0\right)\right\} \right\} ,
\end{equation}
\begin{equation}
\Gamma_{\uparrow}\equiv\frac{1}{2}\mathrm{Re}\left\{ \int_{0}^{\infty}d\tau e^{-i\omega_{d}\tau}\mathrm{Tr}_{B}\left\{ \rho_{B}\hat{\Sigma}_{+}\left(\tau\right)\hat{\Sigma}_{-}\left(0\right)\right\} \right\} ,
\end{equation}
\begin{equation}
\Gamma^{z}\equiv\frac{1}{2}\mathrm{Re}\left\{ \int_{0}^{\infty}d\tau\mathrm{Tr}_{B}\left\{ \rho_{B}\hat{\Sigma}_{z}\left(\tau\right)\hat{\Sigma}_{z}\left(0\right)\right\} \right\} ,
\end{equation}
\begin{equation}
\Gamma_{\pm}^{z}\equiv\frac{1}{2}\mathrm{Re}\left\{ \int_{0}^{\infty}d\tau e^{\pm i\omega\tau}\mathrm{Tr}_{B}\left\{ \rho_{B}\hat{\Sigma}_{z}\left(\tau\right)\hat{\Sigma}_{z}\left(0\right)\right\} \right\} .
\end{equation}
\end{subequations}

\section{Calculation of the decay rates due to tunneling electrons}

\label{ratesApp}

In order to extract the rates that appear in the master equations
\eqref{eq:BlochN}--\eqref{eq:BlochA}, \eqref{eq:generalA}, it
is necessary to calculate correlation functions of the reservoir electronic
operators, for example, 
\begin{widetext}
\noindent 
\begin{align}
\left\langle \hat{\Sigma}_{-}\left(\tau\right)\hat{\Sigma}_{+}\left(0\right)\right\rangle  & =\left(\alpha_{zx}\cos\psi+\frac{\alpha_{\perp}}{2}\sin\psi\right)^{2}\sum_{k,q,\sigma}\left[\left\langle c_{k\sigma\ell}^{\dagger}\left(\tau\right)c_{k\sigma\ell}\left(0\right)\right\rangle \left\langle c_{q\sigma\ell'}\left(\tau\right)c_{q\sigma\ell'}^{\dagger}\left(0\right)\right\rangle +\ell\leftrightarrow\ell'\right]\nonumber \\
 & +\left(\frac{\alpha_{\perp}}{2}\left(\cos\psi-1\right)-\alpha_{zx}\sin\psi\right)^{2}\sum_{k,q}\left[\left\langle c_{k\uparrow\ell}^{\dagger}\left(\tau\right)c_{k\uparrow\ell}\left(0\right)\right\rangle \left\langle c_{q\downarrow\ell'}\left(\tau\right)c_{q\downarrow\ell'}^{\dagger}\left(0\right)\right\rangle +\ell\leftrightarrow\ell'\right]\nonumber \\
 & +\left(\frac{\alpha_{\perp}}{2}\left(\cos\psi+1\right)-\alpha_{zx}\sin\psi\right)^{2}\sum_{k,q}\left[\left\langle c_{k\downarrow\ell}^{\dagger}\left(\tau\right)c_{k\downarrow\ell}\left(0\right)\right\rangle \left\langle c_{q\uparrow\ell'}\left(\tau\right)c_{q\uparrow\ell'}^{\dagger}\left(0\right)\right\rangle +\ell\leftrightarrow\ell'\right].
\end{align}
\end{widetext}

Let us illustrate the calculation of the last product of correlation
functions appearing in the above expression, defining 
\begin{equation}
C\left(\tau\right)\equiv\sum_{k,q}\left\langle c_{q\downarrow\ell'}^{\dagger}\left(\tau\right)c_{q\downarrow\ell'}\left(0\right)\right\rangle \left\langle c_{k\uparrow\ell}\left(\tau\right)c_{k\uparrow\ell}^{\dagger}\left(0\right)\right\rangle .
\end{equation}
Each of these correlations can be expressed in terms of the Fermi-Dirac
distribution, 
\begin{equation}
\left\langle c_{q\downarrow\ell'}^{\dagger}\left(\tau\right)c_{q\downarrow\ell'}\left(0\right)\right\rangle =f_{\ell',\downarrow}\left(\epsilon_{q}\right)e^{i\epsilon_{q}\tau},\label{eq:FDequi1}
\end{equation}
\begin{equation}
\left\langle c_{k\uparrow\ell}\left(\tau\right)c_{k\uparrow\ell}^{\dagger}\left(0\right)\right\rangle =\left[1-f_{\ell,\uparrow}\left(\epsilon_{k}\right)\right]e^{-i\epsilon_{k}\tau},\label{eq:FDequi2}
\end{equation}
with the subscripts of $f$ indicating the lead and spin direction.
Allowing an additional finite voltage $V$ between the leads (may
be set to zero for inter-lead tunneling), we find 
\begin{align}
C\left(\tau\right) & =\sum_{k,q}f_{\ell',\downarrow}\left(\epsilon_{q}\right)\left[1-f_{\ell,\uparrow}\left(\epsilon_{k}\right)\right]e^{-i\left(\epsilon_{k}-\epsilon_{q}+V\right)\tau}\nonumber \\
 & \approx\nu_{\ell'\downarrow}\nu_{\ell\uparrow}\int d\epsilon'\int d\epsilon f\left(\epsilon'\right)\left[1-f\left(\epsilon\right)\right]e^{-i\left(\epsilon-\epsilon'+V\right)\tau},
\end{align}

\noindent where $\nu_{\ell'\downarrow}$ represents the density of
states for $\downarrow$-electrons in lead $\ell'$, and $\nu_{\ell\uparrow}$
the density of states for $\uparrow$-electrons in lead $\ell$. The
densities of states are approximated to be roughly constant near the
Fermi energy.

Next, we Laplace transform this correlation function in order to retrieve
its spectral features. Since we only use the real part of the bath
correlation functions in our analysis (as they represent the rates
governing the master equation; the imaginary parts correspond to shifts
of the subsystem Hamiltonian, whose effect we have verified to be
small), we exploit the relation between the Fourier and Laplace transforms
$\mathcal{F}\left\{ C\right\} =2\mathrm{Re}\left\{ \mathcal{L}\left\{ C\right\} \right\} $
(due to the property $C\left(-\tau\right)=C^{*}\left(\tau\right)$)
and calculate the Fourier transform instead, 
\begin{align}
C\left(\omega\right) & \equiv\int_{-\infty}^{\infty}d\tau e^{i\omega\tau}C\left(\tau\right)\nonumber \\
 & =\nu_{l\downarrow}\nu_{r\uparrow}\int d\epsilon f\left(\epsilon\right)\left[1-f\left(\epsilon+\omega-V\right)\right].
\end{align}

\noindent Let us now define 
\begin{align}
I_{\pm}\left(\omega,V\right) & \equiv\int d\epsilon f\left(\epsilon\right)\left[1-f\left(\epsilon+\omega\pm V\right)\right]\nonumber \\
 & =T\eta_{-}\left(\omega\pm V\right),
\end{align}
with $\eta_{\pm}\left(x\right)\equiv\pm\frac{x}{e^{\pm x}-1}$. We
arrive at the full expressions for the spectral functions,
\begin{widetext}
\begin{subequations} 
\begin{eqnarray}
\left\langle \hat{\Sigma}_{-}\hat{\Sigma}_{+}\right\rangle \left(\omega\right) & = & \left(\alpha_{zx}\cos\psi+\frac{\alpha_{\perp}}{2}\sin\psi\right)^{2}\left(\nu_{\ell\uparrow}\nu_{\ell'\uparrow}+\nu_{\ell\downarrow}\nu_{\ell'\downarrow}\right)\left(I_{-}\left(\omega,V\right)+I_{+}\left(\omega,V\right)\right)\nonumber \\
 &  & +\left(\frac{\alpha_{\perp}}{2}\left(\cos\psi-1\right)-\alpha_{zx}\sin\psi\right)^{2}\left(\nu_{\ell\uparrow}\nu_{\ell'\downarrow}I_{-}\left(\omega,V\right)+\nu_{\ell\downarrow}\nu_{\ell'\uparrow}I_{+}\left(\omega,V\right)\right)\nonumber \\
 &  & +\left(\frac{\alpha_{\perp}}{2}\left(\cos\psi+1\right)-\alpha_{zx}\sin\psi\right)^{2}\left(\nu_{\ell\uparrow}\nu_{\ell'\downarrow}I_{+}\left(\omega,V\right)+\nu_{\ell\downarrow}\nu_{\ell'\uparrow}I_{-}\left(\omega,V\right)\right),
\end{eqnarray}
\begin{eqnarray}
\left\langle \hat{\Sigma}_{+}\hat{\Sigma}_{-}\right\rangle \left(\omega\right) & = & \left(\alpha_{zx}\cos\psi+\frac{\alpha_{\perp}}{2}\sin\psi\right)^{2}\left(\nu_{\ell\uparrow}\nu_{\ell'\uparrow}+\nu_{\ell\downarrow}\nu_{\ell'\downarrow}\right)\left(I_{-}\left(\omega,V\right)+I_{+}\left(\omega,V\right)\right)\nonumber \\
 &  & +\left(\frac{\alpha_{\perp}}{2}\left(\cos\psi-1\right)-\alpha_{zx}\sin\psi\right)^{2}\left(\nu_{\ell\uparrow}\nu_{\ell'\downarrow}I_{+}\left(\omega,V\right)+\nu_{\ell\downarrow}\nu_{\ell'\uparrow}I_{-}\left(\omega,V\right)\right)\nonumber \\
 &  & +\left(\frac{\alpha_{\perp}}{2}\left(\cos\psi+1\right)-\alpha_{zx}\sin\psi\right)^{2}\left(\nu_{\ell\uparrow}\nu_{\ell'\downarrow}I_{-}\left(\omega,V\right)+\nu_{\ell\downarrow}\nu_{\ell'\uparrow}I_{+}\left(\omega,V\right)\right),
\end{eqnarray}
\begin{eqnarray}
\left\langle \hat{\Sigma}_{z}\hat{\Sigma}_{z}\right\rangle \left(\omega\right) & = & \left(\alpha_{zz}\cos\psi+\frac{\alpha_{xz}}{2}\sin\psi\right)^{2}\left(\nu_{\ell\uparrow}\nu_{\ell'\uparrow}+\nu_{\ell\downarrow}\nu_{\ell'\downarrow}\right)\left(I_{-}\left(\omega,V\right)+I_{+}\left(\omega,V\right)\right)\nonumber \\
 &  & +\left(\frac{\alpha_{xz}}{2}\cos\psi-\alpha_{zz}\sin\psi\right)^{2}\left(\nu_{\ell\uparrow}\nu_{\ell'\downarrow}+\nu_{\ell\downarrow}\nu_{\ell'\uparrow}\right)\left(I_{-}\left(\omega,V\right)+I_{+}\left(\omega,V\right)\right).
\end{eqnarray}
\end{subequations}

We may now calculate the contributions of different tunneling processes
to the master equation rates: tip-atom-substrate tunneling {[}$\left(\ell,\ell'\right)=\left(t,s\right)${]},
tip-atom-tip tunneling {[}$\left(\ell,\ell'\right)=\left(t,t\right)${]},
and substrate-atom-substrate tunneling {[}$\left(\ell,\ell'\right)=\left(s,s\right)${]}.
This is done by using the densities of state 
\begin{equation}
\nu_{t\sigma}=\frac{\nu_{t}}{2}\left(1+\sigma p\right),\,\,\,\,\,\nu_{s\uparrow}=\nu_{s\downarrow}=\frac{\nu_{s}}{2},\label{eq:nrup}
\end{equation}
with $p\in\left[0,1\right]$ determining the level of polarization
in the tip. Collecting the different terms together with the proper
coupling constants for each process, we find 
\begin{align}
\Gamma_{\downarrow} & =\frac{1}{2}\left(J_{s}^{2}\nu_{s}^{2}+J_{t}^{2}\nu_{t}^{2}\right)\frac{\omega_{d}}{1-e^{-\beta\omega_{d}}}\left(\alpha_{zx}^{2}+\frac{\alpha_{\perp}^{2}}{2}\right)\nonumber \\
 & +J_{t}^{2}\frac{\nu_{t}^{2}}{2}\frac{\omega_{d}}{1-e^{-\beta\omega_{d}}}p^{2}\left[\left(\alpha_{zx}^{2}-\frac{\alpha_{xx}^{2}}{4}\right)\cos2\psi-\frac{\alpha_{\perp}^{2}}{4}+\alpha_{zx}\alpha_{\perp}\sin2\psi\right]\nonumber \\
 & +J_{s}J_{t}\frac{\nu_{s}\nu_{t}}{2}\frac{V+\omega_{d}}{1-e^{-\beta\left(V+\omega_{d}\right)}}\left[\alpha_{zx}^{2}+\frac{\alpha_{\perp}^{2}}{2}-p\alpha_{\perp}\left(\alpha_{zx}\sin\psi-\frac{\alpha_{\perp}}{2}\cos\psi\right)\right],\label{eq:gammdown}
\end{align}
\begin{align}
\Gamma_{\uparrow} & =\frac{1}{2}\left(J_{s}^{2}\nu_{s}^{2}+J_{t}^{2}\nu_{t}^{2}\right)\frac{\omega_{d}}{e^{\beta\omega_{d}}-1}\left(\alpha_{zx}^{2}+\frac{\alpha_{\perp}^{2}}{2}\right)\nonumber \\
 & +J_{t}^{2}\frac{\nu_{t}^{2}}{2}\frac{\omega_{d}}{e^{\beta\omega_{d}}-1}p^{2}\left[\left(\alpha_{zx}^{2}-\frac{\alpha_{xx}^{2}}{4}\right)\cos2\psi-\frac{\alpha_{\perp}^{2}}{4}+\alpha_{zx}\alpha_{\perp}\sin2\psi\right]\nonumber \\
 & +J_{s}J_{t}\frac{\nu_{s}\nu_{t}}{2}\frac{V-\omega_{d}}{1-e^{-\beta\left(V-\omega_{d}\right)}}\left[\alpha_{zx}^{2}+\frac{\alpha_{\perp}^{2}}{2}+p\alpha_{\perp}\left(\alpha_{zx}\sin\psi-\frac{\alpha_{\perp}}{2}\cos\psi\right)\right],\label{eq:gammup}
\end{align}
\begin{align}
\Gamma^{z} & =J_{s}^{2}\frac{\nu_{s}^{2}}{2}T\left(\alpha_{zz}^{2}+\frac{\alpha_{xz}^{2}}{4}\right)\nonumber \\
 & +J_{t}^{2}\frac{\nu_{t}^{2}}{2}T\left[\alpha_{zz}^{2}\left(1+p^{2}\cos2\psi\right)+\frac{\alpha_{xz}^{2}}{4}\left(1-p^{2}\cos2\psi\right)+p^{2}\alpha_{zz}\alpha_{xz}\sin2\psi\right]\nonumber \\
 & +J_{s}J_{t}V\frac{\nu_{s}\nu_{t}}{2}\left(\alpha_{zz}^{2}+\frac{\alpha_{xz}^{2}}{4}\right),
\end{align}
\begin{align}
\Gamma_{\pm}^{z} & =J_{s}^{2}\frac{\nu_{s}^{2}}{2}T\left(1\pm\frac{\omega}{T}\right)\left(\alpha_{zz}^{2}+\frac{\alpha_{xz}^{2}}{4}\right)\nonumber \\
 & +J_{t}^{2}\frac{\nu_{t}^{2}}{2}T\left(1\pm\frac{\omega}{T}\right)\left[\alpha_{zz}^{2}\left(1+p^{2}\cos2\psi\right)+\frac{\alpha_{xz}^{2}}{4}\left(1-p^{2}\cos2\psi\right)+p^{2}\alpha_{zz}\alpha_{xz}\sin2\psi\right]\nonumber \\
 & +J_{s}J_{t}V\frac{\nu_{s}\nu_{t}}{2}\left(\alpha_{zz}^{2}+\frac{\alpha_{xz}^{2}}{4}\right),\label{eq:gammpmz}
\end{align}
\end{widetext}

where we used the fact that $V\gg T,\omega_{d},\omega$ and $\omega\ll T$,
to make some simplifications. We now employ some additional approximations,
compliant with the experimental setup:
\begin{enumerate}
\item $\alpha_{zz}\gg\alpha_{xz}$ , since according to Eq. \eqref{eq:alphaMatrix-1}
there is a four-orders-of-magnitude difference between them.
\item $p\approx1$ -- taking the polarization level to be maximal allows
us to simplify the above expressions greatly, without modifying the
underlying physics in any meaningful way, as we have explicitly checked.
\item $\psi\approx\frac{\pi}{2}$ -- in accordance with the experiments
\citep{STMexperiment,Willke2018}.
\end{enumerate}
After defining $\tilde{J}_{s}\equiv J_{s}\nu_{s}$, $\tilde{J_{t}}\equiv J_{t}\nu_{t}$,
$\tilde{J_{t}}\equiv r\tilde{J_{s}}\frac{T}{V}$ and $\alpha_{zx}\equiv\chi\alpha_{\perp}$,
we may finally write 
\begin{align}
\Gamma_{\downarrow} & \approx\frac{\tilde{J_{s}}^{2}}{4}\alpha_{\perp}^{2}\left(2\chi^{2}+1+r^{2}\frac{T^{2}}{V^{2}}\right)\frac{\omega_{d}}{1-e^{-\beta\omega_{d}}}\nonumber \\
 & +\frac{\tilde{J_{s}}^{2}}{4}\alpha_{\perp}^{2}\left(2\chi^{2}-2\chi+1\right)\frac{rT\left(1+\frac{\omega_{d}}{V}\right)}{1-e^{-\beta\left(V+\omega_{d}\right)}},\label{eq:kFinals}
\end{align}
\begin{align}
\Gamma_{\uparrow} & \approx\frac{\tilde{J_{s}}^{2}}{4}\alpha_{\perp}^{2}\left(2\chi^{2}+1+r^{2}\frac{T^{2}}{V^{2}}\right)\frac{\omega_{d}}{e^{\beta\omega_{d}}-1}\nonumber \\
 & +\frac{\tilde{J_{s}}^{2}}{4}\alpha_{\perp}^{2}\left(2\chi^{2}+2\chi+1\right)\frac{rT\left(1-\frac{\omega_{d}}{V}\right)}{1-e^{-\beta\left(V-\omega_{d}\right)}},\label{eq:keFinals}
\end{align}
\begin{align}
\Gamma^{z} & \approx\frac{\tilde{J_{s}}^{2}T}{2}\left(1+r\right)\alpha_{zz}^{2},
\end{align}
\begin{align}
\Gamma_{\pm}^{z} & \approx\Gamma^{z}\pm\frac{\tilde{J_{s}}^{2}}{2}\alpha_{zz}^{2}\omega.\label{eq:DeltaInds}
\end{align}
Neglecting terms such as $\frac{\omega_{d}}{V}\approx0.02$ and $\frac{T}{V}\approx0.01$
due to the overwhelming size of the voltage energy scale yields the
expressions that we use in the main text, Eqs. \eqref{eq:kFinal}--\eqref{eq:DeltaInd}.
Note that we assume that $r$ is an order $\mathcal{O}\left(1\right)$
parameter, implying $\tilde{J}_{s}\gg\tilde{J}_{t}$, which is physically
sensible since the adatom is much closer to the substrate than to
the tip.

 \bibliographystyle{apsrev4-1}
\bibliography{\string"C:/Users/galshavi/Desktop/EprStmPrl/answers to PRL round 1/MoshesAnswer/references\string"}

\end{document}